\documentstyle[preprint,aps]{revtex}
\renewcommand{\theequation}{\arabic{section}.\arabic{equation}}
\def\ie{ {i.e.,} }
\def\eg{ {\it e.g.,} }
\def\en{\epsilon_n}
\def\enp{\epsilon_{n'}}
\def\d{{\rm d}}

\def\dt{{\delta \theta}}
\def\o{{\omega_n}}
\def\dt2{(\delta \theta)^2}

\def\p{\sigma}
\def\al{\alpha}
\def\be{\beta}

\def\de{\delta}

\def\im{{\rm i}}

\def\ch{\chi}

\def\lan{\left\langle}
\def\ran{\right\rangle}

\def\ekp{\epsilon_{kp}}
\def\delx{\partial_x}
\def\delxd{\partial_{x'}}

\def\nonum{\nonumber}

\def\dt{\frac{\partial}{\partial T}}

\def\e{{\rm e}}

\def\titheta{\tilde{\theta}}
\def\tiA{\tilde{A}}
\def\tiphi{\tilde{\phi}}
\def\bpsi{\bar{\psi}}
\def\brho{\bar{\rho}}
\def\bn{\bar{N}}
\def\sgn{{\rm sgn}}
\def\Re{{\rm Re}}
\def\lsim{\lower -0.3ex \hbox{$<$} \kern -0.75em \lower 0.7ex \hbox{$\sim$}}
\def\gsim{\lower -0.3ex \hbox{$>$} \kern -0.75em \lower 0.7ex \hbox{$\sim$}}
\def\yen{Y \kern -1.077em =}
\def\virg{\;\;,}
\def\point{\;\,.}
\def\jo #1#2#3#4{#1 {\bf #2}, #4  (#3)}  

\def\PRB{Phys.\ Rev.\ B}
\def\PRL{Phys.\ Rev.\ Lett.}

\def\SSC{Solid State Commun.}

\def\JPC{J.\ Phys.\ C}

\def\JPSJ{J.\ Phys.\ Soc.\ Jpn.}

\def\PTP{Prog.\ Theor.\ Phys.}

\def\ADV{Adv.\ Phys.}

\def\SL{JETP\ Lett.}

\def\PHC{Physica C}

\def\PLA{Phys.\ Lett.\ A}
\def\JLP{J.\ Low Temp.\ Phys.}

\def\IJMPB{Int.\ J.\ Mod.\ Phys.\ B}
\def\SM{Synth.\ Met.}
\def\JMP{J.\ Math.\ Phys.}
\begin{document}
\draft
\title
{ 
Properties of Fluctuations \\
in Two Coupled Chains of Luttinger Liquids 
}
\author
{
Hideo {\sc YOSHIOKA}\thanks{h44770a@nucc.cc.nagoya-u.ac.jp}
and Yoshikazu {\sc SUZUMURA}\thanks{e43428a@nucc.cc.nagoya-u.ac.jp}
}
\address
{
Department of Physics, Nagoya University,\\
Nagoya 464-01, Japan
}
\date{Received 1 July 1996}
\maketitle
\begin{abstract}
 Two chains of Luttinger liquids coupled by both interchain hopping 
 and interchain interaction are investigated. 
 A degeneracy  of  two kinds  of 
 dominant states with in phase and out of phase ordering in the absence of the hopping
 is removed 
 since  the transverse fluctuation of charge density is completely 
gapful and that of spin density has two kinds of excitation with and without gap.  
 The crucial effect of  the interchain hopping 
 on the electronic properties is studied by calculating  
 susceptibilities.
\end{abstract}
\pacs{}
\narrowtext
\section{Introduction}
\setcounter{equation}{0}

It is  well known that  the ground state  
of one-dimensional interacting electron systems is non-Fermi liquid. 
The state, which is  called as ``Luttinger liquid'', is characterized by a 
separation of  spin and charge degrees of freedom, 
 and anomalous exponents of correlation functions which  
 depend on the interaction\cite{Solyom-review,Emery-review,Haldane-review,Fukuyama-Takayama}. 
The parallel chains of the interacting electrons coupled by the interchain hopping 
are basic models of quasi-one-dimensional electron systems. 
Theoretical understanding of these systems  is fundamental for 
 studying electronic properties of  organic conductors
\cite{Ishiguro-Yamaji}, high temperature superconductors
\cite{Andersonconjecture} 
and interacting electron systems applied to strong magnetic field 
(Fractional Quantum Hall Effects) 
\cite{WenFQHE}. 
In addition, the problem is important as a first step toward 
 studying  two-dimensional interacting electron systems.   
 Consequently, there has been a growing interest in the system of the 
coupled chains, in particular  two chains. 
 Several work have been  devoted to investigating  the two chains 
coupled by the interchain hopping for the case of spinless 
Fermion\cite{Wen,Kusmartsev-Luther-Nersesyan,Yakovenko,Yoshioka-Suzumura-1,Yoshioka-Suzumura-2} 
and for the case including spin degree of 
freedom\cite{Castellani-Dicastro-Metzner,Fabrizio,Finkelstein-Larkin,Clarke-Strong-Anderson,Yamaji-Shimoi,Shimoi-Yamaji-Yanagisawa,Nagaosa-Oshikawa,Schulz,Balents-Fisher,Yoshioka-Suzumura-IV}. 
In addition, Anderson localization in such a system has been 
studied based on the above investigation\cite{Kimura-Kuroki-Aoki,Orignac-Giamarchi}.

However, interchain interaction also plays  a role of the interchain 
coupling\cite{Nersesyan-Luther-Kusmartsev,Yoshioka-Suzumura-3}. 
Therefore, by  extending  the work by Finkel'stein and 
Larkin\cite{Finkelstein-Larkin} and that by Schulz\cite{Schulz}, 
 we  study the two chains of Luttinger liquids in the presence of 
  both the interchain hopping and the interchain interaction.  
To our knowledge, much is not known 
  about such a problem in the case 
 including spin degree of freedom. 
We clarify the electronic properties originated from 
 the interchain hopping 
 by calculating excitation spectrum, phase diagram and 
susceptibilities.   
 The gap appears in 
 the excitation spectrum  of the transverse fluctuations. 
 This leads to splitting of 
 the degenerated  states 
 in the absence of the hopping, \ie 
``in phase'' ordering states  and 
 ``out of phase'' ordering states between the chains. 
 From the calculation of charge and spin susceptibilities 
 with $q_x$  and $q_y$ being  
 the longitudinal and transverse wavenumber,  
  it is shown that both  susceptibilities with $q_y = 0$ 
  are the same as those in the absence of the hopping, and   those with $q_y = \pi$ 
show remarkable dependence on $q_x$.  

The plan of the paper is as follows.  
In II, the Hamiltonian of the model is given and is expressed by 
use of the phase variables.  
In III, we study the excitation spectrum 
 and the phase diagram at $T=0$ with $T$ being temperature. 
The charge and the spin susceptibilities are also calculated.  
IV is devoted to discussion on the present results. 
    
\section{Model and Phase Representation}
\setcounter{equation}{0}
\subsection{Model Hamiltonian}
We investigate the system where two  chains  of Luttinger liquids 
are coupled by both the interchain hopping and the interchain interaction. 
 The Hamiltonian is given by 
\begin{equation} \label{eqn:Hamiltonian}
 {\cal H} 
 = {\cal H_{\rm k}} +  {\cal H_{\rm int}}  + {\cal H'_{\rm int}} 
                                                       \virg  
\end{equation}
where  
\begin{eqnarray}
{\cal H_{\rm k}} &=& 
\sum_{k,p,\p} \ekp \left\{ a^{\dagger}_{k,p,\p,1} a_{k,p,\p,1} + ( 1 \to 2 
) \right\}
 - t \sum_{k,p,\p} \left\{ a^{\dagger}_{k,p,\p,1} a_{k,p,\p,2} + ( 1 
\leftrightarrow 2 ) \right\}, 
\label{eqn:2.1}\\
{\cal H_{\rm int}} &=&
{ {\pi v_F g_{2}} \over {L} } \sum_{p,\p,\p'} \sum_{k_1, k_2, q}
\left\{ 
a^{\dagger}_{k_1,p,\p,1} a^{\dagger}_{k_2,-p,\p',1} a_{k_2+q,-p,\p',1} 
a_{k_1-q,p,\p,1} + ( 1 \to 2 )
\right\},
\label{eqn:2.2}\\ 
{\cal H'_{\rm int}} &=&
{ {\pi v_F g'_{2}} \over {L} } \sum_{p,\p,\p'} \sum_{k_1, k_2, q}
\left\{ 
a^{\dagger}_{k_1,p,\p,1} a^{\dagger}_{k_2,-p,\p',2} a_{k_2+q,-p,\p',2} 
a_{k_1-q,p,\p,1} + ( 1 \leftrightarrow 2 ) 
\right\}. 
\label{eqn:2.3} 
\end{eqnarray}
 Equations (\ref{eqn:2.1}), (\ref{eqn:2.2}) and (\ref{eqn:2.3}) 
  express the kinetic energy, the intrachain 
interaction and the interchain interaction, respectively. 
 Quantities $k$ and $\ekp ( = v_F(pk-k_F)) $ denote 
 the momentum and the kinetic energy of a Fermion where 
 $v_F$, $p = +(-)$ and  $k_F$ are 
  the Fermi velocity, the right-going (left-going) 
 state of a Fermion and the Fermi momentum.     
The operator $a^{\dagger}_{k,p,\p,i}$ expresses   
 creation of the Fermion with $k$, $p$, $\p$ and $i$ 
 where $\p = +(-)$ and  $i(=1,2) $ are 
the spin $\uparrow(\downarrow)$ state and the index of the chains. 
The interchain hopping  and the length of the chain are defined  by $t$ and $L$, respectively. 
The normalized quantities,  
$g_{2}$ ($g'_{2}$) denotes 
 the matrix element of the interaction for the intrachain 
(interchain) forward scattering between particles 
  moving oppositely.  
Note that the conventional definition of the elements are given by 
$g \to g/(2\pi v_F)$\cite{Solyom-review}.    

\subsection{Phase Representation}
We represent the above Hamiltonian, 
Eqs.(\ref{eqn:2.1}) $\sim$ (\ref{eqn:2.3})
 and the field operators of Fermions in 
terms of phase variables based on bosonization method. 
  The separation of the Fermi wavenumber 
 due to the hopping is taken into account by use of 
  the unitary transformation, 
$c_{k,p,\p,\mu} = ( - \mu a_{k,p,\p,1} + a_{k,p,\p,2} ) / \sqrt{2}$ 
($\mu = \pm$). 
 Then the kinetic term, ${\cal H}_{\rm k}$ is rewritten as 
\begin{equation}
{\cal H_{\rm k}} = \sum_{k,p,\p,\mu} 
v_F(pk-k_{F\mu}) c^{\dagger}_{k,p,\p,\mu} c_{k,p,\p,\mu},
\label{eqn:2.4}
\end{equation}
 where $k_{F\mu} = k_F - \mu t/v_F$. 
The interaction terms are rewritten as follows,
\begin{eqnarray}
{\cal H}_{\rm int} &+& {\cal H}'_{\rm int} = 
{\cal H}^1_{\rm int} + {\cal H}^2_{\rm int} + {\cal H}^3_{\rm int}, 
\label{eqn:2.5}\\
{\cal H}^1_{\rm int} &=& \frac{\pi v_F}{2L} (g_2 + g_2') 
\sum_{p,\p,\p',\mu,\mu'} \sum_{q}
  \rho_{p,\p,\mu}(q) \rho_{-p,\p',\mu'}(-q) , 
\label{eqn:2.6}\\
{\cal H}^2_{\rm int} &=& \frac{\pi v_F}{2} (g_2 - g_2')
\sum_{p,\p,\p',\mu} \int \d x
\left\{ 
\psi^\dagger_{p,\p,\mu} \psi^\dagger_{-p,\p',\mu} 
\psi_{-p,\p',-\mu} \psi_{p,\p,-\mu}
\right\},  
\label{eqn:2.7}\\
{\cal H}^3_{\rm int} &=& \frac{\pi v_F}{2} (g_2 - g'_2)
\sum_{p,\p,\p',\mu} \int \d x
\left\{
\psi^\dagger_{p,\p,\mu} \psi^\dagger_{-p,\p',-\mu} 
\psi_{-p,\p',\mu} \psi_{p,\p,-\mu}
\right\}, 
\label{eqn:2.8} 
\end{eqnarray}
where 
 $\psi_{p,\p,\mu} \equiv (1\sqrt{L}) \sum_k \e^{\im k x} c_{k,p,\p,\mu}$. 
   The operator  $\rho_{p,\p,\mu}(q) (= \sum_k c^\dagger_{k+q,p,\p,\mu} c_{k,p,\p,\mu})$  
   expresses the density fluctuation 
 around the new Fermi point $k_{F\mu}$ and satisfies   
 the commutation relations, 
$\left[ \rho_{p,\p,\mu}(-q), \rho_{p',\p',\mu'}(q') \right] = \de_{\p \p'} 
\de_{p p'}\de_{q q'} \de_{\mu,\mu'} pqL/(2\pi)$. 
 The Hamiltonian   ${\cal H}^1_{\rm int}$ denotes the interaction 
 described in terms  of the quadratic form of the density fluctuation. 
 For the processes of    Eqs.(\ref{eqn:2.7}) and (\ref{eqn:2.8}),
  the band index is not conserved in ${\cal H}^2_{\rm int}$ and 
 the index is exchanged in  ${\cal H}^3_{\rm int}$,  respectively.  

 Here we define phase variables $\theta_{\pm}(x)$, $\phi_{\pm}(x)$,
$\titheta_{\pm}(x)$ and $\tiphi_{\pm}(x)$ 
as\cite{Yoshioka-Suzumura-3,Suzumura_P,correspondence} 
\begin{eqnarray}
\theta_{\pm}(x) &=&  
 -  \sum_{q \not= 0 } \frac{\pi \im}{\sqrt{2} q L} \e^{(- \al |q|/2 + \im 
q x)} 
\sum_{\p,\mu} \left(  \rho_{+,\p,\mu}(-q)   \pm   \rho_{-,\p,\mu}(-q) 
\right), 
\label{eqn:2.9} \\
\phi_{\pm}(x) &=&  
-  \sum_{q \not= 0 } \frac{\pi \im}{\sqrt{2} q L} \e^{(- \al |q|/2 + \im q 
x)} 
\sum_{\p,\mu} \p \left(  \rho_{+,\p,\mu}(-q)   \pm   \rho_{-,\p,\mu}(-q) 
\right), 
\label{eqn:2.10}\\
\titheta_{\pm}(x) &=&  
 -  \sum_{q \not= 0 } \frac{\pi \im}{\sqrt{2} q L} \e^{(- \al |q|/2 + \im 
q x)} 
\sum_{\p,\mu} \mu \left(  \rho_{+,\p,\mu}(-q)   \pm   \rho_{-,\p,\mu}(-q) 
\right), 
\label{eqn:2.11}\\
\tiphi_{\pm}(x) &=&  
-  \sum_{q \not= 0 } \frac{\pi \im}{\sqrt{2} q L} \e^{(- \al |q|/2 + \im q 
x)} 
\sum_{\p,\mu} \p \mu \left(  \rho_{+,\p,\mu}(-q)   \pm   
\rho_{-,\p,\mu}(-q) \right) , 
\label{eqn:2.12} 
\end{eqnarray}
 where 
 $\al^{-1}$ is a cutoff of the large momentum  corresponding to the band width  $v_F\al^{-1}$. 
 Equations (\ref{eqn:2.9}) $\sim$ (\ref{eqn:2.12}) satisfy  
 commutation relations  given by 
  $[ \theta_+ (x),  \theta_- (x') ] = [ \phi_+ (x),  \phi_- (x') ] = 
   [ \titheta_+ (x),  \titheta_- (x') ] = [ \tiphi_+ (x),  \tiphi_- (x') ] 
= {\rm ln} \left\{1 + \im(x-x')\al^{-1}\right\} 
- {\rm ln} \left\{1 - \im(x-x')\al^{-1}\right\} 
\simeq \im \pi {\rm sgn} ( x - x' )$ and zero for the others. 
The variables, $\theta_{\pm}$ and $\phi_{\pm}$ express the fluctuation 
 of the total charge density  and that of the total spin density, 
respectively,  while 
 $\titheta_{\pm}$ and $\tiphi_{\pm}$ express the transverse fluctuation 
of the charge density  and that of the spin density, respectively.  
 Actually these properties are understood from the facts that
$\delx \theta_+ = ( \pi / \sqrt{2} ) \sum_{p,\p, i} 
\psi^\dagger_{p,\p,i} \psi_{p,\p,i} $,  
$\delx \phi_+ = ( \pi / \sqrt{2} ) \sum_{p,\p, i} \p 
\psi^\dagger_{p,\p,i} \psi_{p,\p,i} $,
$ \delx \titheta_+ = ( - \pi / \sqrt{2} ) \sum_{p,\p} 
\{ \psi^\dagger_{p,\p,1} \psi_{p,\p,2} + h.c. \} $ 
and 	
$\delx \tiphi_+ = ( - \pi / \sqrt{2} ) \sum_{p,\p} \p 
\{ \psi^\dagger_{p,\p,1} \psi_{p,\p,2} +  h.c. \} $ 
 where $\psi_{p,\p,i} = (1/\sqrt{L}) \sum_k \e^{\im k x} a_{k,p,\p,i}$. 
In terms of the phase variables,   
the field operator of Fermions, $\psi_{p,\p,\mu} ( x )$, is expressed as 
\cite{Luther-Peschel,Luther-Emery},  
\begin{eqnarray}
 \psi_{p,\p,\mu} ( x ) &=& { 1 \over \sqrt{2 \pi \al} }
 \exp \left[ 
{\rm i} p k_{F\mu} x + \im \Theta_{p,\p,\mu} + \im \pi \Xi_{p, \p, \mu} 
         \right]  \nonum \\
&\equiv& \psi'_{p,\p,\mu} ( x ) \exp \left( \im \pi \Xi_{p, \p, \mu} 
\right),
\label{eqn:2.13} 
\end{eqnarray}
with 
\begin{equation}
\Theta_{p,\p,\mu} = \frac{1}{2\sqrt{2}}
\Big\{ 
  p \theta_+ + \theta_- + \mu (  p \titheta_+ + \titheta_- )
+ \p ( p \phi_+ + \phi_- ) + \p \mu (  p \tiphi_+ + \tiphi_- )
\Big\}.
\label{eqn:2.14}
\end{equation}
 In Eq.(\ref{eqn:2.13}), the phase factor, $\pi \Xi_{p, \p, \mu}$, 
 is added so that the Fermion operators with different indices 
satisfy the anticommutation relation 
\cite{Solyom-review}. 
 The factor $ \Xi_{p, \p, \mu}$ is given by 
 $\Xi_1 =0$ and $\Xi_i = \sum_{j=1}^{i-1} \hat N_j$, $(i = 2 \sim 8)$ 
 with ${\hat N}_i$ being the number operator of the Fermions with indices $i$  
 where  the index $(p,\p,\mu)$ corresponds to  
$(+,+,+) = 1$, $(+,-,+) = 2$, $(+,+,-) = 3$, $(+,-,-) = 4$, 
$(-,+,+) = 5$, $(-,-,+) = 6$, $(-,+,-) = 7$ and $(-,-,-) = 8$,   
respectively. 
 Note that the above choice of $ \Xi_{p, \p, \mu}$ 
  is not unique. 

By substituting Eqs.(\ref{eqn:2.13}) and (\ref{eqn:2.14}) 
into Eqs.(\ref{eqn:2.7}) and (\ref{eqn:2.8}), and by using the fact that 
${\cal H}_{\rm k} = (\pi v_F / L ) \sum_{p,\p,\mu} \sum_{q} \rho_{p,\p,\mu}(q) \rho_{p,\p,\mu}(-q)$ 
\cite{Luther-Peschel,Luther-Emery,Mattis-Lieb},  
the Hamiltonian ${\cal H} = {\cal H}_T + {\cal H}_R$ can be expressed as, 
\begin{eqnarray}
	 {\cal H}_T & = & \frac{v_\rho}{4 \pi} \int \d x
	 \left\{ \frac{1}{\eta_\rho}(\delx \theta_+)^2 + {\eta_\rho}(\delx 
\theta_-)^2 \right\} 
	       + \frac{v_F}{4 \pi} \int \d x \left\{ (\delx \phi_+)^2 + (\delx 
\phi_-)^2 \right\}, 
\label{eqn:2.15}\\
	 {\cal H}_R 
       & = & 
\frac{v_F}{4 \pi} 
 \int \d x
	 \left\{ \tiA_{\theta,+}(\delx \titheta_+)^2 
              + \tiA_{\theta,-}(\delx \titheta_-)^2 
	      + \tiA_{\phi,+}(\delx \tiphi_+)^2 
              + \tiA_{\phi,-}(\delx \tiphi_-)^2\right\}  \nonum \\
	     & + & { { v_{F} (g_{2}-g'_{2}) } \over {\pi\al^2} } \int {\rm d}x 
         \left\{ \cos \sqrt{2} \titheta_- -  \cos (2 q_0 x -\sqrt{2} 
         \titheta_+) \right\} 
         \left\{ \cos \sqrt{2} \tiphi_- +  \cos \sqrt{2} \tiphi_+ \right\},
\label{eqn:2.16}
\end{eqnarray}
where 
$v_\rho = v_F \sqrt{(1 + 2 g_2 + 2 g'_2)(1 - 2 g_2 - 2 g'_2)}$,  
$\eta_\rho = \sqrt{(1 - 2 g_2 - 2 g'_2)/(1 + 2 g_2 + 2 g'_2)}$, 
$\tiA_{\theta,\pm} = \tiA_{\phi,\pm} = 1$ and 
$q_0 = 2t/v_F$.   
 Here 
${\cal H}_T$ expresses fluctuations of total charge and spin 
 densities, whose excitations are given by $v_\rho |k|$ and $v_F |k|$, respectively.  
On the other hand, ${\cal H}_R$, which expresses the transverse fluctuation,  includes the complex nonlinear terms. 
 The nonlinear terms  including 
$\cos \sqrt{2} \titheta_-$ result from ${\cal H}^2_{\rm int}$ and 
those with $\cos (2 q_0 x -\sqrt{2} \titheta_+)$ are due to 
${\cal H}^3_{\rm int}$.  
In deriving Eq.(\ref{eqn:2.16}),  
we chose a  Hilbert space with 
 the even integers for respective numbers   
$N_1 + N_3$, $N_2 + N_4$, $N_5 + N_7$, $N_6 + N_8$, 
$N_1 + N_5$, $N_1 + N_6$ and $N_1 + N_2$
 where $N_i$  is the eigenvalue of  ${\hat N}_{i}$. 
 The negative sign  of $\cos (2 q_0 x -\sqrt{2} \titheta_+)$
 is due to such a choice of the phase factor, $\pi \Xi_{p,\p,\mu}$. 
 It is noticed that the interchain interaction does not change 
  the structure  of the Hamiltonian, \ie  
    the parameters  
   in the presence of $g_2'$  are obtained by rewriting as  
  $g_2 \to g_2 + g_2'$ in ${\cal H}_T$ and 
$g_2 \to g_2 - g_2'$ in ${\cal H}_R$.

The Hamiltonian, ${\cal H}_R$ can be rewritten as 
\begin{eqnarray}
	 {\cal H}_R & = & v_F \int \d x 
	 \left\{ \bpsi_1^\dagger ( -\im \delx ) \bpsi_1 - \bpsi_2^\dagger ( -\im 
\delx ) \bpsi_2 
	       + \bpsi_3^\dagger ( -\im \delx ) \bpsi_3 - \bpsi_4^\dagger ( -\im 
\delx ) \bpsi_4
	 \right\} \nonum \\
	     & + & \pi v_F (g_2 - g_2') \int \d x
	 \left\{ \im \bpsi^\dagger_3 \bpsi^\dagger_4 - \im \bpsi_4 \bpsi_3 
	         - \bpsi^\dagger_4 \bpsi_3 \e^{- \im 2 q_0 x} - \bpsi^\dagger_3 
\bpsi_4 \e^{\im 2 q_0 x}  
	 \right\} \nonum \\
	 & & \hspace{3.4cm} \times
	 \left\{ \im \bpsi^\dagger_1 \bpsi^\dagger_2 - \im \bpsi_2 \bpsi_1 
	         + \bpsi^\dagger_2 \bpsi_1 + \bpsi^\dagger_1 \bpsi_2  
	 \right\},
	 \label{eqn:2.17}            
\end{eqnarray}
 where 
 $\bpsi_{i}$ ($i = 1 \sim 4$)  are  new Fermion fields
   defined by 
\begin{eqnarray}
	\bpsi_1 & = & \frac{1}{\sqrt{ 2 \pi \al }} 
	              \e^{\im \frac{1}{\sqrt{2}}( \tiphi_+ + \tiphi_-)} 
	              \e^{\im \frac{\pi}{2} ( \hat{\bn_1} + \hat{\bn_2} )}
	              \equiv \bpsi_1'
	              \e^{\im \frac{\pi}{2} ( \hat{\bn_1} + \hat{\bn_2} )}, 
	              \label{eqn:2.18}\\
        \bpsi_2 & = & \frac{1}{\sqrt{ 2 \pi \al }} 
	              \e^{-\im \frac{1}{\sqrt{2}}( \tiphi_+ - \tiphi_-)} 
	              \e^{-\im \frac{\pi}{2} ( \hat{\bn_1} + \hat{\bn_2} )}
	              \equiv \bpsi_2'
	              \e^{-\im \frac{\pi}{2} ( \hat{\bn_1} + \hat{\bn_2} )}, 
	              \label{eqn:2.19}\\ 
	\bpsi_3 & = & \frac{1}{\sqrt{ 2 \pi \al }} 
	              \e^{\im \frac{1}{\sqrt{2}}( \titheta_+ + \titheta_-)} 
	              \e^{\im \frac{\pi}{2} ( \hat{\bn_3} + \hat{\bn_4} ) 
	                + \im \pi ( \hat{\bn_1} + \hat{\bn_2} ) }
	              \equiv \bpsi_3'
	              \e^{\im \frac{\pi}{2} ( \hat{\bn_3} + \hat{\bn_4} ) 
	                + \im \pi ( \hat{\bn_1} + \hat{\bn_2} ) }, 
                  \label{eqn:2.20}\\ 
         \bpsi_4 & = & \frac{1}{\sqrt{ 2 \pi \al }} 
	              \e^{-\im \frac{1}{\sqrt{2}}( \titheta_+ - \titheta_-)} 
	              \e^{-\im \frac{\pi}{2} ( \hat{\bn_3} + \hat{\bn_4} ) 
	                + \im \pi ( \hat{\bn_1} + \hat{\bn_2} ) }
	              \equiv \bpsi_4'
	              \e^{-\im \frac{\pi}{2} ( \hat{\bn_3} + \hat{\bn_4} ) 
	                + \im \pi ( \hat{\bn_1} + \hat{\bn_2} ) }   . 
	              \label{eqn:2.21}
\end{eqnarray}
In Eqs.(\ref{eqn:2.18})$\sim$(\ref{eqn:2.21}),  
 the phase factors with the number operator 
 are introduced again to satisfy the anticommutation relation. 
 Equation (\ref{eqn:2.17}) is derived by choosing the Hilbert space with   
 both $\bn_1 + \bn_2 $ and $\bn_3 + \bn_4 $ being even integers. 
 It is noted that 
  the  phase variables in Eqs.(\ref{eqn:2.18}) $\sim$ (\ref{eqn:2.21}) 
 are expressed as 
\begin{eqnarray}
\titheta_{\pm}(x) &=&  
 -  \sum_{q \not= 0 } \frac{\sqrt{2} \pi \im}{q L} \e^{(- \al |q|/2 + \im 
q x)} 
\left(  \brho_3 (-q)   \pm   \brho_4 (-q) \right) \virg 
\label{eqn:2.22}\\
\tiphi_{\pm}(x) &=&  
-  \sum_{q \not= 0 } \frac{\sqrt{2} \pi \im}{q L} \e^{(- \al |q|/2 + \im q 
x)} 
\left(  \brho_1(-q)   \pm   \brho_2 (-q) \right) \virg  
\label{eqn:2.23} 
\end{eqnarray}
 where 
 $\brho_j (q) \equiv \sum_k \bpsi^\dagger_j (k+q) \bpsi_j (k)$ 
with $\bpsi_j = 1/\sqrt{L} \sum_k \bpsi_j (k) \e^{\im k x}$ ($j = 1 \sim 4$).

\section{Properties at Low Temperatures}
\setcounter{equation}{0}
By using renormalization group method, Finkel'stein and Larkin
\cite{Finkelstein-Larkin} 
showed that the nonlinear terms in Eq.(\ref{eqn:2.16}) with $g_2 \neq g_2'$ tend to 
the strong coupling in the limit of low energy and 
 the terms without (with) the misfit parameter, $2 q_0$ 
  become relevant (irrelevant). 
Schulz insisted that the transverse charge excitation is completely gapful and 
that of spin excitation has two kinds of excitation with gap and gapless 
from the symmetry of the Hamiltonian. 
 Here we explicitly show the results by use of 
 the mean field approximation    
in which the terms including the misfit parameter are neglected. 
This method is expected to be effective in the limit of 
 strong coupling  and has an advantage of the straightforward 
 calculation of the several quantities.  
It should be noted that the break of the balance between $\titheta_+$ 
 and $\titheta_-$ due to  the misfit parameter may lead to 
  the renormalization of  $\tilde{\eta}_\rho$ 
 defined by $(\tiA_{\theta,-}/\tiA_{\theta,+})^{1/2}$.  
 We neglect such an effect as zeroth-approximation 
 since the system tends to strong coupling and the gap appears. 
 On the other hand, 
 $\tilde{\eta}_\p ( \equiv (\tiA_{\phi,-}/\tiA_{\phi,+})^{1/2} )$  
  remains unity  due to 
 the balance between $\tiphi_+$ and $\tiphi_-$. 
 The present method is effective for the energy lower than the hopping, and 
 then the large momentum cutoff $\al^{-1}$ must be read as $t/v_F$.  

\subsection{Excitation Spectrum}

 By making use of the mean-field approximation, 
  ${\cal H}_R$ with $g_2 \neq g_2'$ is rewritten as 
\begin{eqnarray}
 {\cal H}_R = {\cal H}^{12}_{\rm MF} + {\cal H}^{34}_{\rm MF}
 - \Delta \Delta' L /(2 \pi v_F (g_2-g'_2)) \virg 
\end{eqnarray}
 where 
\begin{eqnarray}
	{\cal H}^{12}_{\rm MF} & = & v_F \int \d x
	\left\{ 
	\bpsi_1^\dagger ( -\im \delx ) \bpsi_1 - \bpsi_2^\dagger ( -\im \delx ) 
\bpsi_2 \right\} \nonum \\ 
	&+& \frac{\Delta}{2} \int \d x
	\left\{ \im \bpsi^\dagger_1 \bpsi^\dagger_2 - \im \bpsi_2 \bpsi_1 
	         + \bpsi^\dagger_2 \bpsi_1 + \bpsi^\dagger_1 \bpsi_2  \right\}, 
	\label{eqn:3.1}\\
	{\cal H}^{34}_{\rm MF} & = & v_F \int \d x
	\left\{ 
	\bpsi_3^\dagger ( -\im \delx ) \bpsi_3 - \bpsi_4^\dagger ( -\im \delx ) 
\bpsi_4 \right\} 
	 +  \Delta' \int \d x
	\left\{ \im \bpsi^\dagger_3 \bpsi^\dagger_4 - \im \bpsi_4 \bpsi_3 
	\right\}. 
	\label{eqn:3.2} 
\end{eqnarray}
 The quantities, $\Delta$ and $\Delta'$ in Eqs.(\ref{eqn:3.1}) and 
 (\ref{eqn:3.2})  are gap parameters 
 determined by the following self-consistent equations,
\begin{eqnarray}
	\frac{\Delta}{2} & = & \pi v_F ( g_2 - g'_2 ) 
\left\{ \im \lan \bpsi^\dagger_3 \bpsi^\dagger_4 \ran - \im \lan \bpsi_4 
\bpsi_3 \ran \right\}, 
\label{eqn:3.3} \\ 
	\Delta' & = & 	\pi v_F ( g_2 - g'_2) 
\left\{ \im \lan \bpsi^\dagger_1 \bpsi^\dagger_2 \ran - \im \lan \bpsi_2 
\bpsi_1 \ran 
	 + \lan \bpsi^\dagger_2 \bpsi_1 \ran  + \lan \bpsi^\dagger_1 \bpsi_2 \ran 
\right\}. 
\label{eqn:3.4}                      
\end{eqnarray}
The eigenvalues, $\omega_{12}$  of Eq.(\ref{eqn:3.1}) and 
 $\omega_{34}$ of Eq.(\ref{eqn:3.2}) 
 are calculated as 
\begin{eqnarray}
 \omega_{12} &=& 
   \left\{
   \begin{array}{l}
     \pm E_k  \equiv \pm \sqrt{ (v_F k)^2 + \Delta^2 } \virg \\ 
     \pm v_F k \virg 
   \end{array}
   \right. 
   \label{eqn:n3.5}                                        \\ 
 \omega_{34} & =& \pm E'_k  \equiv \pm \sqrt{ (v_F k)^2 + \Delta'^2 } 
\point 
\label{eqn:n3.6} 
\end{eqnarray}
 The excitations of Eqs.(\ref{eqn:n3.5}), 
whose spectral weights are 1/2,  
 are obtained by ${\cal H}^{12}_{\rm MF}$ 
in terms of Majorana Fermion as 
\begin{eqnarray}
{\cal H}_{\rm MF}^{12}
&=& \frac{v_F}{2} \int \d x 
    \left\{ C_1(-\im \delx)C_1 - C_2(-\im \delx)C_2 \right\}
    + \im \Delta \int \d x C_1 C_2 \nonum \\
&+& \frac{v_F}{2} \int \d x 
    \left\{ 
    C_0(-\im \delx)C_0 - C_3(-\im \delx)C_3 
    \right\} \virg 
\label{eqn:3.5} 
\end{eqnarray}
 where 
 $C_0 = \im / \sqrt{2}
( \e^{\im \pi /4} \bpsi_1 - \e^{-\im \pi /4} \bpsi_1^\dagger)$,
$C_1 = 1 / \sqrt{2}
( \e^{\im \pi /4} \bpsi_1 + \e^{-\im \pi /4} \bpsi_1^\dagger)$,
$C_2 = 1 / \sqrt{2}
( \e^{- \im \pi /4} \bpsi_2 + \e^{\im \pi /4} \bpsi_2^\dagger)$ and 
$C_3 = \im / \sqrt{2}
( \e^{-\im \pi /4} \bpsi_2 - \e^{\im \pi /4} \bpsi_2^\dagger)$.  
Finkel'stein and Larkin\cite{Finkelstein-Larkin} 
  have already obtained  the same form 
 as the first line in Eq. (\ref{eqn:3.5})  
 by  replacing  $ \cos \sqrt{2} \titheta_-$ 
 in Eq.(\ref{eqn:2.16}) with a value of a fixed point.   
However they did not show the gapless excitation which contributes 
to low energy properties \eg specific heat\cite{Schulz}.      

 The  gap equations of Eqs.(\ref{eqn:3.3}) and (\ref{eqn:3.4}) 
 are calculated (see  Appendix A)  as 
\begin{eqnarray}
	\Delta & = & - 2 \Delta' ( g_2 - g'_2 ) \log \frac{\xi_c + 
	\sqrt{\xi_c^2 + \Delta'^2}}{|\Delta'|}, 
	\label{eqn:3.6} \\
	\Delta' & = & - \Delta ( g_2 - g'_2 ) \log \frac{\xi_c + 
	\sqrt{\xi_c^2 + \Delta^2}}{|\Delta|}, 
    \label{eqn:3.7} 	
\end{eqnarray}
where $\xi_c$ is cut-off energy of the order of $t$. 
From the the gap equations, it is found   that 
 both $(-\Delta,-\Delta')$ and  $(\Delta,\Delta')$  are solutions,  
 and that  $\sgn ( \Delta  \Delta') = -1$ for $g_2 - g_2' > 0$ 
and $\sgn ( \Delta  \Delta' ) = 1$ for $g_2 - g_2' < 0$.  
 The solutions of Eqs.(\ref{eqn:3.6}) and (\ref{eqn:3.7}) in the case of  $\Delta' > 0$ 
 is shown in Fig.\ref{fig:1}. 

\subsection{Possible States}
We examine phase diagram which shows the most divergent state. 
Since  the dominant contribution 
 in the low energy limit is given by ${\cal H}^2_{\rm int}$ in 
 Eq.(\ref{eqn:2.7}) 
 \cite{Finkelstein-Larkin}, 
 we rewrite the term as 
\begin{eqnarray}
{\cal H}^2_{\rm int} 
& = & 
\frac{\pi v_F}{4} ( g_2 - g'_2 ) \sum_{p,\p,\p'} \int \d x
\Big\{ 
\big(\sum_{\mu'} \psi^\dagger_{-p,\p,\mu'} \psi^\dagger_{p,\p',\mu'} \big)
\big(\sum_\mu \psi_{p,\p',\mu} \psi_{-p,\p,\mu} \big) \nonum \\
& & \hspace{4.5cm} -  
\big(\sum_{\mu'} \mu' \psi^\dagger_{-p,\p,\mu'} \psi^\dagger_{p,\p',\mu'} 
\big)
\big(\sum_\mu \mu \psi_{p,\p',\mu} \psi_{-p,\p,\mu} \big) \Big\} \nonum \\
& = & 
- \frac{\pi v_F}{4} ( g_2 - g'_2 ) \sum_{p,\p,\p'} \int \d x
\Big\{ 
\big(\sum_{\mu'} \psi^\dagger_{-p,\p,-\mu'} \psi_{p,\p',\mu'} \big)
\big(\sum_\mu \psi^\dagger_{p,\p',\mu} \psi_{-p,\p,-\mu} \big) \nonum \\
& & \hspace{4.5cm} -  
\big(\sum_{\mu'} \mu' \psi^\dagger_{-p,\p,-\mu'} \psi_{p,\p',\mu'} \big)
\big(\sum_\mu \mu \psi_{p,\p',\mu} \psi_{-p,\p,-\mu} \big) \Big\} 
                                         \point
\label{eqn:3.8}  
\end{eqnarray}
Since the  states should be selected so as to gain 
 the  energy from Eq.(\ref{eqn:3.8}), 
the possible states 
 in the case from $g_2 - g'_2 > 0$ are given by 
\begin{eqnarray}
S_{-}^{\p,\p'} 
&\equiv& \sum_{\mu} \mu \psi_{p,\p,\mu} \psi_{-p,\p',\mu} 
= - \left\{ \psi_{p,\p,1} \psi_{-p,\p',2} + (1 \leftrightarrow 2) \right\} \nonum \\
& \sim & 
\frac{\im}{\pi \al} 
\e^{\frac{\im}{\sqrt{2}}\theta_-}
\e^{\frac{\im p}{2\sqrt{2}}(\p- \p')\phi_+} 
\e^{\frac{\im}{2\sqrt{2}}(\p + \p')\phi_-} 
 \sin \left\{ \frac{\titheta_-}{\sqrt{2}} + p 
\frac{\p-\p'}{2\sqrt{2}}\tiphi_+ 
+ \frac{\p +\p'}{2\sqrt{2}}\tiphi_- \right\}, 
\label{eqn:3.9} \\
DW_{+}^{\p,\p'} 
&\equiv& \sum_{\mu} \psi^\dagger_{p,\p,\mu} \psi_{-p,\p',-\mu} 
 = - \left\{ \psi^\dagger_{p,\p,1} \psi_{-p,\p',1} - (1 \to 2)  \right\} \nonum \\
& \sim & 
\frac{-\im}{\pi \al}
\e^{-\im 2 p k_F x}  
\e^{\frac{-\im p}{\sqrt{2}}\theta_+}
\e^{\frac{-\im p}{2\sqrt{2}}(\p + \p')\phi_+} 
\e^{\frac{-\im}{2\sqrt{2}}(\p - \p')\phi_-}
 \sin \left\{ \frac{\titheta_-}{\sqrt{2}} + p 
\frac{\p-\p'}{2\sqrt{2}}\tiphi_+ 
+ \frac{\p +\p'}{2\sqrt{2}}\tiphi_- \right\}, \nonum \\ 
& &
\label{eqn:3.10} 
\end{eqnarray} 
and those in the case of  $g_2 - g'_2 < 0$ are given by 
\begin{eqnarray}
S_{+}^{\p,\p'}
&\equiv& \sum_{\mu}  \psi_{p,\p,\mu} \psi_{-p,\p',\mu} 
 =  \psi_{p,\p,1} \psi_{-p,\p',1} + (1 \to 2) \nonum \\
& \sim & 
\frac{1}{\pi \al} 
\e^{\frac{\im}{\sqrt{2}}\theta_-}
\e^{\frac{\im p}{2\sqrt{2}}(\p- \p')\phi_+} 
\e^{\frac{\im}{2\sqrt{2}}(\p + \p')\phi_-} 
\cos \left\{ \frac{\titheta_-}{\sqrt{2}} + p 
\frac{\p-\p'}{2\sqrt{2}}\tiphi_+ 
+ \frac{\p +\p'}{2\sqrt{2}}\tiphi_- \right\}, 
\label{eqn:3.11} \\
DW_{-}^{\p,\p'}
&\equiv& \sum_{\mu} \mu \psi^\dagger_{p,\p,\mu} \psi_{-p,\p',-\mu} 
 =  - \left\{ \psi^\dagger_{p,\p,1} \psi_{-p,\p',2} - (1 \leftrightarrow 
2) \right\} \nonum \\
& \sim & 
\frac{1}{\pi \al}
\e^{-\im 2 p k_F x}  
\e^{\frac{-\im p}{\sqrt{2}}\theta_+}
\e^{\frac{-\im p}{2\sqrt{2}}(\p + \p')\phi_+} 
\e^{\frac{-\im}{2\sqrt{2}}(\p - \p')\phi_-} 
\cos \left\{ \frac{\titheta_-}{\sqrt{2}} + p 
\frac{\p-\p'}{2\sqrt{2}}\tiphi_+ 
+ \frac{\p +\p'}{2\sqrt{2}}\tiphi_- \right\}. \nonum \\
& &
\label{eqn:3.12} 
\end{eqnarray} 
Here $DW^{\p,\p'}_{-}$ (  $DW^{\p,\p'}_{+}$ )
 expresses 
density wave with interchain and out of phase ordering  
( with intrachain and out of phase ordering), 
 while $S^{\p,\p'}_{-}$ ( $S^{\p,\p'}_{+}$) expresses 
 superconductivity with interchain and in phase ordering  
 ( with  intrachain and in phase ordering).  
 The most dominant state between  
$S_-^{\p,\p'}$ and $DW_+^{\p,\p'}$ for $g_2 > g_2'$ 
($S_+^{\p,\p'}$ and $DW_-^{\p,\p'}$ for $g_2 < g_2'$ )
 is determined  by  
 the total charge and spin fluctuations. 
 By noting that  the correlation functions for 
  the total fluctuations are calculated as, 
\begin{eqnarray}
\lan
\e^{\frac{-\im}{\sqrt{2}}\theta_-(x)}
\e^{\frac{-\im p}{2\sqrt{2}}(\p- \p')\phi_+(x)} 
\e^{\frac{-\im}{2\sqrt{2}}(\p + \p')\phi_-(x)}
\e^{\frac{\im}{\sqrt{2}}\theta_-(0)}
\e^{\frac{\im p}{2\sqrt{2}}(\p- \p')\phi_+(0)} 
\e^{\frac{\im}{2\sqrt{2}}(\p + \p')\phi_-(0)}
\ran
&\sim&
\left( \frac{\al}{|x|} \right)^{\frac{1}{2} + \frac{1}{2\eta_\rho} },
\nonum \\
& & 
\label{eqn:3.13} \\
\lan
\e^{\frac{\im p}{\sqrt{2}}\theta_+(x)}
\e^{\frac{\im p}{2\sqrt{2}}(\p + \p')\phi_+(x)} 
\e^{\frac{\im}{2\sqrt{2}}(\p - \p')\phi_-(x)}
\e^{\frac{-\im p}{\sqrt{2}}\theta_+(0)}
\e^{\frac{-\im p}{2\sqrt{2}}(\p + \p')\phi_+(0)} 
\e^{\frac{-\im}{2\sqrt{2}}(\p - \p')\phi_-(0)}
\ran
&\sim&
\left( \frac{\al}{|x|} \right)^{\frac{1}{2} + \frac{\eta_\rho}{2} }, 
\nonum \\
\label{eqn:3.14}
\end{eqnarray}
 we obtain the phase diagram 
  in Fig.\ref{fig:2}.  
 The exponent  of the correlation function of CDW  
 is the same as that of SDW, and the exponent of 
 singlet superconductivity is the same as that of
  triplet superconductivity, respectively. 
 This comes from the symmetry of the interaction 
  for the  spin degree of freedom, 
 as  is also seen  in spin dependent Tomonaga model 
  with the isotropic interaction 
 \cite{Fukuyama-Takayama}.    
Therefore the phase diagram shown in Fig.\ref{fig:2} is essentially the same as that of 
spinless Fermion studied previously\cite{Yoshioka-Suzumura-3}. 
In the repulsive case of $g_2 + g_2' >0$, the most dominant state is the density wave. 
However, the interchain hopping leads to 
  the superconducting state being  subdominant even in such a region.

\subsection{Susceptibilities}
In this subsection, we calculate 
 the charge  susceptibilities, $ \ch_\rho(q_x, q_y; \im \o)$,  and 
 the spin susceptibilities, 
 $\ch_\p (q_x, q_y; \im \o)$, 
 in the case of  $q_x \ll 2k_F$  and   $q_y = 0$ or $\pi$ 
 which are defined  as 
\begin{equation}
\ch_\nu (q_x, q_y; \im \o) 
=  \int^\be_0 \d \tau \int \d (x - x') \e^{\im \o \tau} \e^{-\im q_x (x - x') } 
	 \ch_\nu (x-x', q_y; \tau), 
	 \label{eqn:3.15} 
\end{equation}
and $\nu$ = $\rho$ or $\p$.  
In Eq.(\ref{eqn:3.15}), 
\begin{eqnarray}
	\ch_\rho(x-x', q_y; \tau) 
	 & = & 
	 \frac{1}{2}\lan T_\tau 
	 \left\{ \rho(x, 1 ; \tau) + \e^{\im q_y} \rho(x, 2 ; \tau) \right\}
	 \left\{ \rho(x', 1 ; 0) + \e^{\im q_y} \rho(x', 2 ; 0) \right\}
	 \ran, 
	 \label{eqn:3.16} \\
    \ch_\p (x-x', q_y; \tau) 
	 & = & 
	 \frac{1}{2}\lan T_\tau 
	 \left\{ m(x, 1 ; \tau) + \e^{\im q_y} m(x, 2 ; \tau) \right\}
	 \left\{ m(x', 1 ; 0) + \e^{\im q_y} m(x', 2 ; 0) \right\} 
	 \ran,  
	 \label{eqn:3.17} 
\end{eqnarray}
 where 
$\rho(x,i;\tau)  = \sum_{p,\p} \psi^\dagger_{p,\p,i}(x;\tau) 
\psi_{p,\p,i}(x;\tau)$ and 
$m(x,i;\tau)  = \sum_{p,\p} \p \psi^\dagger_{p,\p,i}(x;\tau) 
\psi_{p,\p,i}(x;\tau)$ denote 
 operators of the charge and spin densities  
at the  $i$-th chain, respectively.  

At first
we consider the case of $q_y = 0$. 
From 
Eq.(\ref{eqn:2.15}), both 
$\ch_\rho(x-x', 0; \tau)$ and $\ch_\p(x-x', 0; \tau)$ are  
calculated as
\begin{eqnarray}
\ch_\rho(x-x', 0; \tau) 
&=&  
\frac{1}{\pi^2} \lan T_\tau \delx \theta_+(x,\tau) \delxd \theta_+(x',0) 
\ran \nonum \\
&=&
\frac{2 \eta_\rho}{\pi v_\rho}
\frac{1}{\be L} \sum_{q_x} \frac{(v_\rho q_x)^2}{\o^2 + (v_\rho q_x)^2} 
\e^{\im q_x(x-x') - \im \o \tau}, 
\label{eqn:3.18} \\
\ch_\p(x-x', 0; \tau) 
&=&
\frac{1}{\pi^2} 
\lan T_\tau \delx \phi_+(x,\tau) \delxd \phi_+(x',0) \ran \nonum \\
&=&
\frac{2}{\pi v_F}
\frac{1}{\be L} \sum_{q_x} \frac{(v_F q_x)^2}{\o^2 + (v_F q_x)^2} 
\e^{\im q_x(x-x') - \im \o \tau} .
\label{eqn:3.19} 
\end{eqnarray}
Therefore ($\im \omega_n \to \omega$), one obtains  
\begin{eqnarray}
{\rm Re} \ch_\rho(q_x, 0;\omega) 
& = &
\frac{2 \eta_\rho}{\pi v_\rho}
\frac{(v_\rho q_x)^2}{(v_\rho q_x)^2 - \omega^2} \virg  
\label{eqn:3.20} \\
{\rm Re} \ch_\p(q_x, 0;\omega)
& = &
\frac{2}{\pi v_F} 
\frac{(v_F q_x)^2}{(v_F q_x)^2 - \omega^2} \virg 
\label{eqn:3.21} 
\end{eqnarray}
 which are  familiar to  Luttinger 
liquid\cite{Solyom-review}. 

Next we examine  the case of $q_y = \pi$,  
 for  which  Eqs.(\ref{eqn:3.16}) and (\ref{eqn:3.17}) are expressed as 
\begin{eqnarray}
& & \ch_\rho (x-x', \pi; \tau) \nonum \\
	 & = &  
	 \frac{1}{2}\lan T_\tau 
	 \Bigg( 
	 \sum_{p',\p',\mu'} \psi^\dagger_{p',\p',-\mu'}(x;\tau) 
\psi_{p',\p',\mu'}(x;\tau)
	 \Bigg)
	 \Bigg( 
	 \sum_{p,\p,\mu} \psi^\dagger_{p,\p,\mu}(x') \psi_{p,\p,-\mu}(x')
	 \Bigg) \ran,
	 \label{eqn:3.22} \\
& & \ch_\p (x-x', \pi; \tau) \nonum \\ 
	 & = &  
	 \frac{1}{2} \lan T_\tau 
	 \Bigg( 
	 \sum_{p',\p',\mu'} \p' \psi^\dagger_{p',\p',-\mu'}(x;\tau) 
\psi_{p',\p',\mu'}(x;\tau)
	 \Bigg)
	 \Bigg( 
	 \sum_{p,\p,\mu} \p \psi^\dagger_{p,\p,\mu}(x') \psi_{p,\p,-\mu}(x')
	 \Bigg) \ran. \nonum \\
	 & &
	 \label{eqn:3.23} 
\end{eqnarray}
After some manipulations (Appendix B), the static susceptibilities, 
${\rm Re} \ch_\rho(q_x,\pi;0)$ and ${\rm Re} \ch_\p(q_x,\pi;0)$ at $T=0$ are 
respectively obtained as 
\begin{eqnarray}
{\rm Re} \ch_\rho(q_x,\pi;0)
 & = & 
\frac{1}{2\pi} \int \d k
\Bigg\{ 
\frac{1}{E_k + E'_{k - q_x - q_0}}
\Big( 
1 - \frac{\xi_k}{E_k} \frac{\xi_{k - q_x -q_0}}{E'_{k - q_x -q_0}}
+ \frac{\Delta}{E_k} \frac{\Delta'}{E'_{k - q_x -q_0}}
\Big) 
\nonum \\
& & \hspace{1cm}  + 
\frac{1}{E_k + E'_{k + q_x - q_0}}
\Big( 
1 - \frac{\xi_k}{E_k} \frac{\xi_{k + q_x -q_0}}{E'_{k + q_x -q_0}}
+ \frac{\Delta}{E_k} \frac{\Delta'}{E'_{k + q_x -q_0}}
\Big)
\Bigg\},  
            \label{eqn:3.24}           \\
{\rm Re} \ch_\p(q_x,\pi,0)
&= &  
\frac{1}{2\pi} 
\Bigg\{ 
 P \int^{q_x + q_0} \d k \left( 1 + \frac{\xi_k}{E'_k} \right)
\frac{1}{E'_k - \xi_k + v_F(q_x + q_0)} \nonum \\
& &\hspace{0.5cm} + 
 P \int_{q_x + q_0} \d k \left( 1 - \frac{\xi_k}{E'_k} \right)
\frac{1}{E'_k + \xi_k - v_F(q_x + q_0)} \nonum \\
& &\hspace{0.5cm} + 
 P \int^{-q_x + q_0} \d k \left( 1 + \frac{\xi_k}{E'_k} \right)
\frac{1}{E'_k - \xi_k + v_F(-q_x + q_0)} \nonum \\
& &\hspace{0.5cm} + 
 P \int_{-q_x + q_0} \d k \left( 1 - \frac{\xi_k}{E'_k} \right)
\frac{1}{E'_k + \xi_k - v_F(-q_x + q_0)} 
\Bigg\} \nonum \\
&= & 
\frac{1}{\pi v_F}
\left\{ 
2 - \frac{\Delta'^2}{2 \epsilon_+^2} \ln \left( 1 + \frac{\epsilon_+^2}{\Delta'^2} \right) 
  - \frac{\Delta'^2}{2 \epsilon_-^2} \ln \left( 1 + \frac{\epsilon_-^2}{\Delta'^2} \right) 
\right\},
\label{eqn:3.25}
\end{eqnarray}
where $P$ denotes the principal value,  
 $ \xi_k = v_{\rm F}k$ and $\epsilon_\pm = v_F (\pm q_x + q_0)$. 
 In Fig.\ref{fig:3} and Fig.\ref{fig:4}, we show 
 the normalized quantities 
 ${\bar \ch}_\rho(q_x,\pi;0)$ and 
 ${\bar \ch}_\rho(q_x,\pi;0)$ which are defined by 
 ${\rm Re} \ch_\rho (q_x,\pi;0) / (2/\pi v_F)$ 
  and  
${\rm Re} \ch_\p (q_x,\pi,0)/(2/\pi v_F)$, respectively. 
The cutoff energy, $\xi_c$, defined in Eqs.(\ref{eqn:3.6}) 
and (\ref{eqn:3.7}) is taken as $2t$. 
Note that Eqs.(\ref{eqn:3.24}) and (\ref{eqn:3.25}) are valid in the case 
of $|q_x \pm q_0| \lsim q_0$. 

Equation (\ref{eqn:3.24}) shows that the value of ${\rm Re}\ch_\rho(q_x,\pi;0)$ 
in the case of $g_2 - g_2' < 0$ is larger than that 
in the case of $g_2 - g_2' > 0$ because
$g_2 - g_2' < 0$ ($> 0$) leads to $\sgn( \Delta  \Delta') = 1$ ($-1$).  
Such dependence on $g_2 - g_2'$ is also found in the 
absence of the hopping (see Eq.(\ref{eqn:D15})). 
 On the other hand, ${\rm Re} \ch_\p (q_x,\pi;0)$ is independent of sign 
 of the relative interaction. 
 This  result seems to correspond to the fact  that 
${\rm Re} \ch_\p (q_x,\pi;0)$ in the absence of the hopping 
is independent of $g_2 - g_2'$ (see Eq.(\ref{eqn:D16})). 

 In Fig.\ref{fig:3},  
   $\Re \ch_\rho(q_x,\pi;0)$ takes  
   a minimum  around $q_x = q_0$   in the case of  $g_2 - g_2'>0$ 
 and 
 has a small  dependence on $q_x$ and $g_2-g_2'$, \ie
 being  nearly unity for $g_2 - g_2'<0$. 
 On the other hand,  $\Re \ch_{\sigma}(q_x,\pi;0)$  in Fig.\ref{fig:4} 
 takes the minimum for both $g_2 - g_2'>0$ and $g_2 - g_2'<0$. 
 The characteristic dependence is due to the separation of Fermi wavenumber
 \ie $k_{F-} - k_{F+} = q_0$ and 
 the gap in the transverse fluctuation, 
 both of which result from the interchain hopping. 
Note that the $q_x$-dependence  of these 
 $\Re \ch_\rho(q_x,\pi;0)$ and $\Re \ch_\sigma(q_x,\pi;0)$
  does not change qualitatively by the choice of   $\xi_c/t$.


\section{Discussion}
\setcounter{equation}{0}
In the present paper, 
 we studied 
  the low temperature properties of  two chains 
  coupled by the interchain hopping 
 and the interchain interaction where 
 the interactions  of only  the forward scattering 
 between oppositely moving particles  were taken into 
 account as a simplest  model of Luttinger liquid. 

 There are 
  four kinds of excitations  originated from the fluctuations of 
 total charge, total spin, transverse charge and  transverse spin 
 respectively. 
 The total fluctuations 
 which show the gapless excitation 
 are the same as those in the 
absence of the interchain hopping (Appendix C). 
 On the other hand, the transverse fluctuations of both charge density and spin density 
 which are  expressed by the complicated  non-linear terms 
  are crucial in the presence of the interchain hopping.  
 By utilizing the mean field approximation, it was shown that 
 the transverse fluctuation of the charge is completely gapful and 
 that of the spin has the two kinds of excitations with and without gap. 
 
The most dominant states are obtained  as 
$DW^{\p,\p'}_+$ for $g_2 > |g_2'|$, 
$DW^{\p,\p'}_-$ for $g_2' > |g_2|$, 
$S^{\p,\p'}_+$ for $g_2 < -|g_2'|$ and 
$S^{\p,\p'}_-$ for $g_2' <- |g_2|$, respectively 
 where   density wave (superconductivity) belongs to  
out of phase (in phase) ordering between the chains, \ie 
  the transverse wavenumber being  $\pi$($0$). 
 We note that 
 in the case of  quasi one-dimensional electron system 
  with  only the hopping of pairs\cite{Suzumura-Fukuyama}, 
   the density wave with the transverse wave number $(\pi,\pi)$ or 
superconductivity with $(0,0)$ has maximum critical temperature. 
 It is worth while noting that the states of the superconductivity remain 
subdominant even for the repulsive  interaction, \ie 
$g_2 + g'_2 > 0$. 
Such a result may be the important point toward understanding of 
 the competition of 
 superconductivity and SDW observed in quasi one-dimensional conductors, $\rm{(TMTSF)_2 
X}$\cite{Yoshioka-Suzumura-IV}.

 Possible states in the absence of the interchain hopping 
 are obtained by calculating  correlation functions for  
 $S^{\p,\p'}_{||}$ 
 $(= \psi_{p,\p,1} \psi_{-p,\p',1}$ 
 or $\psi_{p,\p,2} \psi_{-p,\p',2})$,  
 $S^{\p,\p'}_{\bot}$
 $( = \psi_{p,\p,1} \psi_{-p,\p',2}$ 
 or $\psi_{p,\p,2} \psi_{-p,\p',1})$, 
 $DW^{\p,\p'}_{||}$ 
 $( = \psi^\dagger_{p,\p,1} \psi_{-p,\p',1}$
 or $\psi^\dagger_{p,\p,2} \psi_{-p,\p',2})$ 
 and  
 $DW^{\p,\p'}_{\bot}$ 
 $(= \psi^\dagger_{p,\p,1} \psi_{-p,\p',2}$ 
 or $\psi^\dagger_{p,\p,2} \psi_{-p,\p',1}) $ 
 which express the order parameters 
 of the intrachain superconductivity, the interchain 
 superconductivity, the intrachain density wave and the interchain density wave, 
 respectively (Appendix C). 
 The phase diagram in the absence of the interchain hopping 
 is shown in  Fig.\ref{fig:5}. 
 By comparing the phase diagram in Fig.\ref{fig:2} and that in Fig.\ref{fig:5}, 
 it is found  that 
  the energy gain due to the  
 interchain hopping removes  the degeneracy  
  of ``in phase'' and ``out of phase'' ordering.

 The $q_x$-dependence of 
 charge and spin susceptibilities were calculated 
 for both  $q_y = 0$ and $q_y = \pi$. 
 The susceptibilities  in the case of $q_y = 0$ 
 are   the same  as those  in the absence of the hopping 
 since  the total dynamics is not affected by the hopping. 
 On the other hand, 
    the static susceptibilities, 
 $\Re \ch_\rho(q_x,\pi;0)$ for $g_2 -  g_2'>0$ and $\Re \ch_\p (q_x,\pi;0)$   
 in the case of  $q_y = \pi$ 
 show  the minimum  around $q_x = q_0$   
  which is ascribed to  the separation of the Fermi wavenumber and the 
 excitation gaps of the  transverse fluctuation 
  in the presence of  the  interchain hopping.     
  The fact that  $\Re \ch_\rho(q_x,\pi;0)$ 
 in the case of  $g_2 -g'_2<0 $  
 is larger than that  in the case of  $g_2 -g'_2>0 $  
 is  found  also in the absence of the interchain hopping. 

We treated the system with the interaction processes of the forward 
scattering between the oppositely 
moving particles as the simplest model of Luttinger liquid.  
However, the two chains coupled by the interchain hopping have 
been known to show the various 
properties depending on the parameters of the system.   
The repulsive backward scattering becomes relevant in the low energy limit and 
opens the gap in the excitations of the total spin, and thus modifies the 
electronic properties\cite{Schulz}. 
This fact is different from the strictly one-dimensional case,  
where the repulsive backward scattering is renormalized to zero. 
The two chains of Hubbard Model shows the richer phases 
depending on the magnitude of the intrachain interaction, 
the interchain hopping and the filling\cite{Fabrizio,Balents-Fisher}. 
In addition, it has been reported that the two chains coupled by 
both the Coulomb repulsion and the exchange interaction show the superconductivity
\cite{Shelton-Tsvelik}. 
Therefore further  investigations are needed 
 to identify the ground state of two 
coupled chains as a crossover from one 
dimension to higher dimension.

\section*{Acknowledgment}

The authors would like to thank A. M. Finkel'stein for discussion.   
This work was financially supported  by the 
Grant-in-Aid for Scientific  Research on the priority area,   
Novel Electronic States in Molecular Conductors, from the 
Ministry of Education.

\newpage

\appendix
\renewcommand{\theequation}{\Alph{section}\arabic{equation}}
\section{Green functions of Eqs.(3.2) and (3.3)}

 We calculate  Green functions of Eqs.(\ref{eqn:3.1})  and (\ref{eqn:3.1}). 
 By using the solutions of Majorana Fermions which are   
calculated from Eq.(\ref{eqn:3.5}), 
 the Green functions corresponding to the 
transverse spin fluctuations are calculated as 
\begin{eqnarray}
- \lan T_\tau \bpsi_1(x,\tau) \bpsi_1^\dagger (x',\tau') \ran 
&=& \frac{1}{2 \be L} \sum_{k,\en}
\e^{\im k(x-x') - \im \en (\tau - \tau') } 
\left\{ 
\frac{1}{\im \en - \xi_k} 
+ \frac{\im \en + \xi_k}{(\im \en + E_k)(\im \en - E_k)} 
\right\}, 
\label{eqn:A1} \\ 
- \lan T_\tau \bpsi_2(x,\tau) \bpsi_2^\dagger (x',\tau') \ran 
&=& \frac{1}{2 \be L} \sum_{k,\en}
\e^{\im k(x-x') - \im \en (\tau - \tau') } 
\left\{ 
\frac{1}{\im \en + \xi_k} 
+ \frac{\im \en - \xi_k}{(\im \en + E_k)(\im \en - E_k)} 
\right\}, 
\label{eqn:A2} \\ 
- \lan T_\tau \bpsi_1(x,\tau) \bpsi_1 (x',\tau') \ran 
&=& \frac{\im}{2 \be L} \sum_{k,\en}
\e^{\im k(x-x') - \im \en (\tau - \tau') } 
\left\{ 
\frac{1}{\im \en - \xi_k} 
- \frac{\im \en + \xi_k}{(\im \en + E_k)(\im \en - E_k)} 
\right\}, 
\label{eqn:A3} \\ 
- \lan T_\tau \bpsi_2(x,\tau) \bpsi_2 (x',\tau') \ran 
&=& \frac{- \im}{2 \be L} \sum_{k,\en}
\e^{\im k(x-x') - \im \en (\tau - \tau') } 
\left\{ 
\frac{1}{\im \en + \xi_k} 
- \frac{\im \en - \xi_k}{(\im \en + E_k)(\im \en - E_k)} 
\right\}, 
\label{eqn:A4} \\ 
- \lan T_\tau \bpsi_1(x,\tau) \bpsi_2 (x',\tau') \ran
&=& \frac{\im}{2 \be L} \sum_{k,\en}
\e^{\im k(x-x') - \im \en (\tau - \tau') }
\frac{\Delta}{(\im \en + E_k)(\im \en - E_k)} \virg 
\label{eqn:A5} \\
- \lan T_\tau \bpsi_1(x,\tau) \bpsi_2^\dagger (x',\tau') \ran
&=& \frac{1}{2 \be L} \sum_{k,\en}
\e^{\im k(x-x') - \im \en (\tau - \tau') }
\frac{\Delta}{(\im \en + E_k)(\im \en - E_k)} \virg 
\label{eqn:A6}  
\end{eqnarray} 
 where $T_\tau$ is the time ordering operator,  
 $\xi_k = v_F k$, $1/\beta = T$ and $\en = (2n+1) \pi T$.  
On the other hand, 
by diagonalizing Eq.(\ref{eqn:3.2}), the Green functions expressing the 
transverse charge fluctuations are calculated as 
\begin{eqnarray}
- \lan T_\tau \bpsi_3(x,\tau) \bpsi_3^\dagger (x',\tau') \ran 
&=& \frac{1}{\be L} \sum_{k,\en}
\e^{\im k(x-x') - \im \en (\tau - \tau') }  
\frac{\im \en + \xi_k}{(\im \en + E'_k)(\im \en - E'_k)} \virg  
\label{eqn:A7} \\ 
- \lan T_\tau \bpsi_4(x,\tau) \bpsi_4^\dagger (x',\tau') \ran 
&=& \frac{1}{\be L} \sum_{k,\en}
\e^{\im k(x-x') - \im \en (\tau - \tau') }  
\frac{\im \en - \xi_k}{(\im \en + E'_k)(\im \en - E'_k)} \virg  
\label{eqn:A8} \\ 
- \lan T_\tau \bpsi_3(x,\tau) \bpsi_4 (x',\tau') \ran
&=& \frac{\im}{\be L} \sum_{k,\en}
\e^{\im k(x-x') - \im \en (\tau - \tau') }
\frac{\Delta'}{(\im \en + E'_k)(\im \en - E'_k)} \point 
\label{eqn:A9} 
\end{eqnarray} 
From these Green functions, one obtains in the limit of 
 $T=0$,
\begin{eqnarray}
   \lan \bpsi_1^\dagger \bpsi_2^\dagger \ran 
   & = &- \lan \bpsi_2 \bpsi_1 \ran 
          = - \im \lan \bpsi_2^\dagger \bpsi_1 \ran
              = - \im \lan \bpsi_1^\dagger \bpsi_2 \ran
                                 \nonumber \\
       & = &\{ \im \Delta / (4 \pi v_F) \}   
     \log \{ (\xi_c + \sqrt{\xi_c^2 + \Delta^2} ) / |\Delta| \},  
                                 \label{eqn:A10}     \\ 
 \lan \bpsi_3^\dagger \bpsi_4^\dagger \ran
   & = & - \lan \bpsi_4 \bpsi_3 \ran 
       = \{ \im \Delta'/ (2 \pi v_F) \} 
      \log \{ ( \xi_c + \sqrt{\xi_c^2 + \Delta'^2} ) / |\Delta'| \} 
                             \label{eqn:A11}            \point 
\end{eqnarray}
  By substituting Eqs.(\ref{eqn:A10}) and (\ref{eqn:A11})
  into  Eqs.(\ref{eqn:3.3}) and (\ref{eqn:3.4}), 
  one obtains Eqs.(\ref{eqn:3.6}) and 
(\ref{eqn:3.7}). 


\section{Calculation of susceptibilities, Eqs. (3.27) and 
(3.28)}

Neglecting the constant whose absolute value is unity, 
 the quantity $ F_{\rho (\p)} \equiv \sum_{p,\p,\mu} (\p) \psi^\dagger_{p,\p,\mu} 
\psi_{p,\p,-\mu}$
is expressed as, 
\begin{eqnarray}
F_{\rho (\p)} &=& 
     \sum_{p,\p,\mu} p \mu (\p) \psi'^\dagger_{p,\p,\mu} \psi'_{p,\p,-\mu} 
                  \nonumber \\
 & = & \frac{1}{2 \pi \al} \sum_{p,\p,\mu}  p \mu (\p) \e^{\im p \mu q_0 x} 
      \exp \left\{ \frac{- \im p \mu}{\sqrt{2}} ( \titheta_+ + p 
\titheta_- ) \right\}
      \exp \left\{ \frac{- \im p \mu \p }{\sqrt{2}} ( \tiphi_+ + p 
\tiphi_- ) \right\} 
                       \nonumber \\
 & = & \e^{\im q_0 x} 
      \left\{ \bpsi'^\dagger_3 \bpsi'^\dagger_1 +(-) \bpsi'^\dagger_3 
\bpsi'_1 
            + \bpsi'_4 \bpsi'_2 +(-) \bpsi'_4 \bpsi'^\dagger_2 
      \right\} \nonum \\ 
 &  &  \hspace{0.5cm} + \e^{- \im q_0 x} 
      \left\{ - \bpsi'_1 \bpsi'_3 -(+) \bpsi'^\dagger_1 \bpsi'_3 
            - \bpsi'^\dagger_2 \bpsi'^\dagger_4 -(+) \bpsi'_2 
\bpsi'^\dagger_4 
      \right\} 
              \nonumber \\
& = & \e^{\im q_0 x} 
      \left\{ \bpsi^\dagger_3 \left[ \bpsi^\dagger_1 -(+) \e^{- \im \pi/2} 
\bpsi_1 \right] 
            - \bpsi_4 \left[ \bpsi_2 -(+) \e^{- \im \pi/2} \bpsi^\dagger_2 
\right] 
      \right\} \nonum \\
 &  &   \hspace{0.5cm}      - \e^{- \im q_0 x} 
      \left\{ \left[ \bpsi_1 -(+) \e^{\im \pi/2} \bpsi^\dagger_1 \right] 
\bpsi_3  
            - \left[ \bpsi^\dagger_2 -(+) \e^{\im \pi/2} \bpsi_2 \right] 
\bpsi^\dagger_4  
      \right\}  \virg
\label{eqn:B1}          
\end{eqnarray}
 which is derived by calculating 
 $\lan T_\tau F^\dagger_{\rho(\p)}(x;\tau) 
F_{\rho(\p)}(x';0) 
\ran$ 
 with the precise treatment of 
  the negative sign originated from the phase 
factors  in terms of  $\hat N_{p,\p,\mu}$ and $\hat {\bar N}_i$ $(i = 1 \sim 4)$. 
 Equation (\ref{eqn:B1})  shows that the susceptibilities with $q_y = \pi$ are 
expressed by only the transverse degree of freedoms.  

 By use of Eqn.(\ref{eqn:A1}) $\sim$ (\ref{eqn:A9}), 
 the quantity  $\lan T_\tau F^\dagger_{\rho(\p)}(x;\tau) F_{\rho(\p)}(x';0) 
\ran$ is calculated as
\begin{eqnarray}
& & \lan T_\tau F^\dagger_{\rho(\p)}(x;\tau) F_{\rho(\p)}(x';0) \ran 
\nonum \\
& = & \e^{- \im q_0 (x - x')} 
\Big\langle 
T_\tau 
\left\{ 
\left[ \bpsi_1(x;\tau) -(+) \e^{\im \pi/2} \bpsi^\dagger_1(x;\tau) \right] 
\bpsi_3(x;\tau)  
- \left[ \bpsi^\dagger_2(x;\tau) -(+) \e^{\im \pi/2} \bpsi_2(x;\tau) 
\right] \bpsi^\dagger_4(x;\tau)  
\right\}
\nonum \\
& & \hspace{2.5cm} \times 
\left\{ \bpsi^\dagger_3(x') \left[ \bpsi^\dagger_1(x') -(+) \e^{- \im 
\pi/2} \bpsi_1(x') \right] 
- \bpsi_4(x') \left[ \bpsi_2(x') -(+) \e^{- \im \pi/2} \bpsi^\dagger_2(x') 
\right] 
\right\} 
\Big\rangle
\nonum \\
& + & ( x \leftrightarrow x', \tau \to -\tau) \nonum \\
& \equiv & \e^{- \im q_0 (x - x')} H_{\rho(\p)}(x-x',\tau) + ( x 
\leftrightarrow x', \tau \to -\tau) \virg 
\label{eqn:B2}  
\end{eqnarray}
 where 
\begin{eqnarray}
 H_{\rho}(x-x',\tau)& &  
 =  
2 \left( \frac{1}{\be L} \right)^2  \sum_{k,k'} \sum_{\en,\enp} 
\e^{\im (k+k')(x-x') - \im (\en + \enp)\tau} \nonum \\
&\times&
\Bigg\{ \left( \frac{1}{\im \enp - E'_{k'}} \frac{1}{\im \en - E_{k}}  
            + \frac{1}{\im \enp + E'_{k'}} \frac{1}{\im \en + E_{k}} 
\right) 
    (u_k u'_{k'} + v_k v'_{k'})^2
                                         \nonum \\
&+&  \left( \frac{1}{\im \enp - E'_{k'}} \frac{1}{\im \en 
+ E_{k}}  
            + \frac{1}{\im \enp + E'_{k'}} \frac{1}{\im \en - E_{k}} 
\right) 
             (u_k v'_{k'} - v_k u'_{k'})^2
\Bigg\}, 
\label{eqn:B3}  \\
 H_{\p}(x-x',\tau) & &  =   2 \left( \frac{1}{\be L} \right)^2 
   \sum_{k,k'} \sum_{\en,\enp} 
 \e^{\im (k+k')(x-x') - \im (\en + \enp)\tau} \nonum \\
&\times& 
    \Bigg\{ 
    \left( \frac{u'^2_k}{\im \en - E'_{k}} 
          + \frac{v'^2_k}{\im \en + E'_{k}} \right)
    \frac{1}{\im \enp - \xi_{k'}}
+              
\left( \frac{v'^2_k}{\im \en - E'_{k}} + \frac{u'^2_k}{\im \en + E'_{k}} 
\right)
\frac{1}{\im \enp + \xi_{k'}}
\Bigg\} \virg  
\label{eqn:B4}  
\end{eqnarray}
 and the factors, $u_k$, $v_k$, $u'_k$, and $v'_k$ are defined by 
\begin{eqnarray}
u_k &=& \sqrt{ \frac{1}{2} ( 1 + \frac{\xi_k}{E_k} ) },
\label{eqn:B5}  \\
v_k &=& \sqrt{ \frac{1}{2} ( 1 - \frac{\xi_k}{E_k} ) } \sgn \Delta,
\label{eqn:B6}  \\ 
u'_k &=& \sqrt{ \frac{1}{2} ( 1 + \frac{\xi_k}{E'_k} ) },
\label{eqn:B7}  \\
v'_k &=& \sqrt{ \frac{1}{2} ( 1 - \frac{\xi_k}{E'_k} ) } \sgn \Delta'
            \point 
\label{eqn:B8}   
\end{eqnarray}

 From Eqs.(\ref{eqn:B3}) and (\ref{eqn:B4}), 
  $\ch_{\rho}(q_x,\pi,\im \o)$ and 
$\ch_{\p}(q_x,\pi,\im \o)$
 are calculated as,
\begin{eqnarray}
 \ch_{\rho}(q_x,\pi,\im \o) 
 & = & \frac{1}{L} \sum_{k}
  \Bigg\{
   \Big(
- \frac{ f(E_k) - f(-E'_{k'}) }{E_k + E'_{k'} - \im \o}
- \frac{ f(-E_k) - f(E'_{k'}) }{-E_k - E'_{k'} - \im \o}
\Big) 
           (u_k u'_{k'} + v_k v'_{k'})^2
\Bigg|_{k' = - k + q_0 + q_x} \nonum \\
&  &  \hspace{1.5cm} +
\Big(
- \frac{ f(-E_k) - f(-E'_{k'}) }{-E_k + E'_{k'} - \im \o}
- \frac{ f(E_k) - f(E'_{k'}) }{E_k - E'_{k'} - \im \o}
\Big) 
         (u_k v'_{k'} - v_k u'_{k'})^2
 \Bigg|_{k' = - k + q_0 + q_x} \nonum \\ 
& & \hspace{3.0cm} + \left( \o \to -\o, q_x \to -q_x \right)
\Bigg\}, 
\label{eqn:B11}   \\
\ch_{\p}(q_x,\pi,\im \o) 
& = &   
 \frac{1}{L} \sum_{k}
\Bigg\{ 
u'^2_k
\Big(
- \frac{f(E'_k) - f(-\xi_{k'})}{E'_k + \xi_{k'} - \im \o}
- \frac{f(-E'_k) - f(\xi_{k'})}{-E'_k - \xi_{k'} - \im \o}
\Big) \Bigg|_{k' = - k + q_0 + q_x} \nonum \\
  & & 
  + v'^2_k
   \Big(
- \frac{f(-E'_k) - f(-\xi_{k'})}{-E'_k + \xi_{k'} - \im \o}
- \frac{f(E'_k) - f(\xi_{k'})}{E'_k - \xi_{k'} - \im \o}
\Big)  \Bigg|_{k' = - k + q_0 + q_x} 
                       \nonumber \\
 & & \hspace{3.0cm}
  + \left( \o \to -\o, q_x \to -q_x \right)
\Bigg\} \virg   
\label{eqn:B12}  
\end{eqnarray} 
where $f(z) = 1/(\e^{\be z} + 1 )$ is a Fermi distribution function. 
 Equations (\ref{eqn:B11}) and (\ref{eqn:B12}) 
in the limit of $T=0$   lead to 
 Eqs.(\ref{eqn:3.24}) and (\ref{eqn:3.25}).


\section{Properties in the absence of the hopping}
In this case, 
 by using density operators, 
$\rho_{p,\p,\nu}(q)( \equiv  \sum_k a^\dagger_{k+q,p,\p,\nu} a_{k,p,\p,\nu})$ with 
 $\nu = +$ and $\nu = -$ being  the chain index $i=1$ and $i=2$  
respectively,  
 Eqs.(\ref{eqn:2.1}) $\sim$ 
(\ref{eqn:2.3}) 
in the absence of the  hopping  
 can be written as,  
\begin{eqnarray}
{\cal H}_{\rm k} & = & \frac{\pi v_F}{L} \sum_{p,\p,\nu} \sum_q 
\rho_{p,\p,\nu}(q) \rho_{p,\p,\nu}(-q), 
\label{eqn:D1}  \\
{\cal H}_{\rm int} & = &  \frac{\pi v_F g_2}{L} \sum_{p,\p,\nu} \sum_q
\left\{ \rho_{p,\p,\nu}(q) \rho_{-p,\p,\nu}(-q) + \rho_{p,\p,\nu}(q) 
\rho_{-p,-\p,\nu}(-q) \right\}, 
\label{eqn:D2} \\
{\cal H}'_{\rm int} & = &  \frac{\pi v_F g'_2}{L} \sum_{p,\p,\nu} \sum_q
\left\{ \rho_{p,\p,\nu}(q) \rho_{-p,\p,-\nu}(-q) + \rho_{p,\p,\nu}(q) 
\rho_{-p,-\p,-\nu}(-q) \right\}.
\label{eqn:D3}
\end{eqnarray}
Then the  Hamiltonian is rewritten in terms of  the phase variables as
\begin{eqnarray}
{\cal H} &=& 
\frac{v_\rho}{4 \pi} \int \d x 
\left\{ \frac{1}{\eta_\rho} (\delx \theta_+)^2 + \eta_\rho (\delx 
\theta_-)^2 \right\}
+ \frac{v'_\rho}{4 \pi} \int \d x 
\left\{ \frac{1}{\eta'_\rho} (\delx \theta'_+)^2 + \eta'_\rho (\delx 
\theta'_-)^2 \right\}
\nonum \\
& + & \frac{v_F}{4 \pi} \int \d x 
\left\{ (\delx \phi_+)^2 +  (\delx \phi_-)^2 + (\delx \phi'_+)^2 +  (\delx 
\phi'_-)^2 \right\} \virg 
\label{eqn:D4}
\end{eqnarray}
where 
\begin{eqnarray}
\theta_{\pm}(x) &=&  
 -  \sum_{q \not= 0 } \frac{\pi \im}{\sqrt{2} q L} \e^{(- \al |q|/2 + \im 
q x)} 
\sum_{\p,\nu} \left(  \rho_{+,\p,\nu}(-q)   \pm   \rho_{-,\p,\nu}(-q) 
\right),  
\label{eqn:D5} \\
\theta'_{\pm}(x) &=&  
 -  \sum_{q \not= 0 } \frac{\pi \im}{\sqrt{2} q L} \e^{(- \al |q|/2 + \im 
q x)} 
\sum_{\p,\nu} \nu \left(  \rho_{+,\p,\nu}(-q)   \pm   \rho_{-,\p,\nu}(-q) 
\right), 
\label{eqn:D6} \\
\phi_{\pm}(x) &=&  
-  \sum_{q \not= 0 } \frac{\pi \im}{\sqrt{2} q L} \e^{(- \al |q|/2 + \im q 
x)} 
\sum_{\p,\nu} \p \left(  \rho_{+,\p,\nu}(-q)   \pm   \rho_{-,\p,\nu}(-q) 
\right), 
\label{eqn:D7} \\
\phi'_{\pm}(x) &=&  
-  \sum_{q \not= 0 } \frac{\pi \im}{\sqrt{2} q L} \e^{(- \al |q|/2 + \im q 
x)} 
\sum_{\p,\nu} \p \nu \left(  \rho_{+,\p,\nu}(-q)   \pm   
\rho_{-,\p,\nu}(-q) \right) \virg 
\label{eqn:D8}
\end{eqnarray}
 and
 $\eta_\rho = \sqrt{(1 - 2g_2 - 2g_2')/(1 + 2g_2 + 2g_2')}$,  
$\eta'_\rho = \sqrt{(1 - 2g_2 + 2g_2')/(1 + 2g_2 - 2g_2')}$, 
$v_\rho = v_F \sqrt{(1 + 2g_2 + 2g_2')(1 - 2g_2 - 2g_2')}$ and
$v'_\rho = v_F \sqrt{(1 + 2g_2 - 2g_2')(1 - 2g_2 + 2g_2')}$.  
 The quantities, $\theta_{\pm}$ and $\phi_{\pm}$ defined here 
 are the same as those of Eqs.(\ref{eqn:2.9}) and (\ref{eqn:2.10}), 
 while  $\theta'_{\pm}$ and $\phi'_{\pm}$ are different from  
$\titheta_{\pm}$ and $\tiphi_{\pm}$ of Eqs.(\ref{eqn:2.11}) and 
(\ref{eqn:2.12}).   
 Since  the parts including $\theta_\pm$ and $\phi_\pm$ 
 in Eqs.(\ref{eqn:D4}) is the same as  Eq.(\ref{eqn:2.15}), 
  the dynamics of total fluctuation does not depend on 
 the interchain hopping.  
 By making use of  field operators, $\psi_{p,\p,\nu}$ 
 expressed as 
\begin{equation}
\psi_{p,\p,\nu} = \frac{\e^{\im p k_F x}}{\sqrt{2 \pi \al}} 
\exp \left[ \frac{\im}{2\sqrt{2}} 
\left\{ p \theta_+ + \theta_- + \nu (p \theta'_+ + \theta'_-) 
  + \p (p \phi_+ + \phi_-) + \p \nu (p \phi'_+ + \phi'_-) \right\}
  \right] \virg  
\label{eqn:D9}  
\end{equation}
 correlation functions of the order parameters are calculated as,
\begin{eqnarray}
\lan S^{\p,\p' \dagger}_{||}(x) S^{\p,\p'}_{||}(0) \ran 
& \sim & 
\left( \frac{\al}{|x|} \right)^{1 + \frac{1}{2}(\frac{1}{\eta_\rho} + 
\frac{1}{\tilde \eta_\rho}) }, 
\label{eqn:D11} \\
\lan S^{\p,\p' \dagger}_{\bot}(x) S^{\p,\p'}_{\bot}(0) \ran 
& \sim & 
\left( \frac{\al}{|x|} \right)^{1 + \frac{1}{2}(\frac{1}{\eta_\rho} + 
{\tilde \eta_\rho}) }, 
\label{eqn:D12} \\
\lan DW^{\p,\p' \dagger}_{||}(x) DW^{\p,\p'}_{||}(0) \ran 
& \sim & 
\left( \frac{\al}{|x|} \right)^{1 + \frac{1}{2}({\eta_\rho} + {\tilde 
\eta_\rho}) }, 
\label{eqn:D13} \\
\lan S^{\p,\p' \dagger}_{\bot}(x) S^{\p,\p'}_{\bot}(0) \ran 
& \sim & 
\left( \frac{\al}{|x|} \right)^{1 + \frac{1}{2}({\eta_\rho} + 
\frac{1}{\tilde \eta_\rho}) }, 
\label{eqn:D14}
\end{eqnarray}
where $S^{\p,\p'}_{||} = \psi_{p,\p,\nu} \psi_{-p,\p',\nu}$,  
$S^{\p,\p'}_{\bot} = \psi_{p,\p,\nu} \psi_{-p,\p',-\nu}$, 
$DW^{\p,\p'}_{||} = \psi^\dagger_{p,\p,\nu} \psi_{-p,\p',\nu}$ and  
$DW^{\p,\p'}_{\bot} = \psi^\dagger_{p,\p,\nu} \psi_{-p,\p',-\nu}$ express 
the order parameters of the intrachain superconductivity, the interchain 
superconductivity, the intrachain density wave and the interchain density 
wave, 
respectively . 
 In Eq.(\ref{eqn:D9}),  we neglect the phase factor given by 
  $\hat N_{p,\p,\nu}$, because 
the particle number of each branch is conserved in the absence of 
 the hopping. 
   The  exponent  of the correlation functions of ``{in phase}''  
ordering is the  same as that of  
 the   ``{out of phase}'' ordering  
 in addition to
{degeneracy} in spin degree of freedom. 
The phase diagram obtained from Eqs.(\ref{eqn:D11}) $\sim$ (\ref{eqn:D14}) is shown in Fig.\ref{fig:5}.

 Susceptibilities are calculated as follows. 
 From Eq.(\ref{eqn:D4}), it is found that 
  ${\rm Re} \ch_\rho(q_x, 0, \omega)$ and ${\rm Re} \ch_\p(q_x, 
0, \omega)$ are the same as Eqs.(\ref{eqn:3.20}) and 
(\ref{eqn:3.21}).  
 By noting that 
\begin{eqnarray*}
\chi_\rho(x-x',\pi,\tau) & = & \frac{1}{\pi^2} 
\lan T_\tau \delx \theta'_+(x,\tau) \delxd \theta'_+(x',0) \ran, \\
\chi_\p(x-x',\pi,\tau) & = & \frac{1}{\pi^2} 
\lan T_\tau \delx \phi'_+(x,\tau) \delxd \phi'_+(x',0) \ran, 
\end{eqnarray*}
 we obtain ${\rm Re} \ch_\rho(q_x, \pi, \omega)$ and ${\rm Re} \ch_\p(q_x, 
\pi, \omega)$ written as  
\begin{eqnarray}
{\rm Re} \ch_\rho(q_x, \pi,\omega) 
& = &
\frac{2 \eta'_\rho}{\pi v'_\rho}
\frac{(v'_\rho q_x)^2}{(v'_\rho q_x)^2 - \omega^2}, 
\label{eqn:D15} \\
{\rm Re} \ch_\p(q_x, \pi,\omega)
& = &
\frac{2}{\pi v_F} 
\frac{(v_F q_x)^2}{(v_F q_x)^2 - \omega^2}, 
\label{eqn:D16}
\end{eqnarray}
which are the forms familiar to the  Luttinger liquid. 
Note that ${\rm Re} \ch_\rho(q_x, \pi,\omega)$ is suppressed (enhanced) in 
the case of $g_2 - g'_2 > 0$ $(<0)$  while 
 ${\rm Re} \ch_\p(q_x, \pi,\omega)$ is independent of 
$g_2 - g'_2$. 
 Thus it turns out that 
the two chains coupled only by the interchain interaction 
 remain as  the Luttinger liquid. 
   


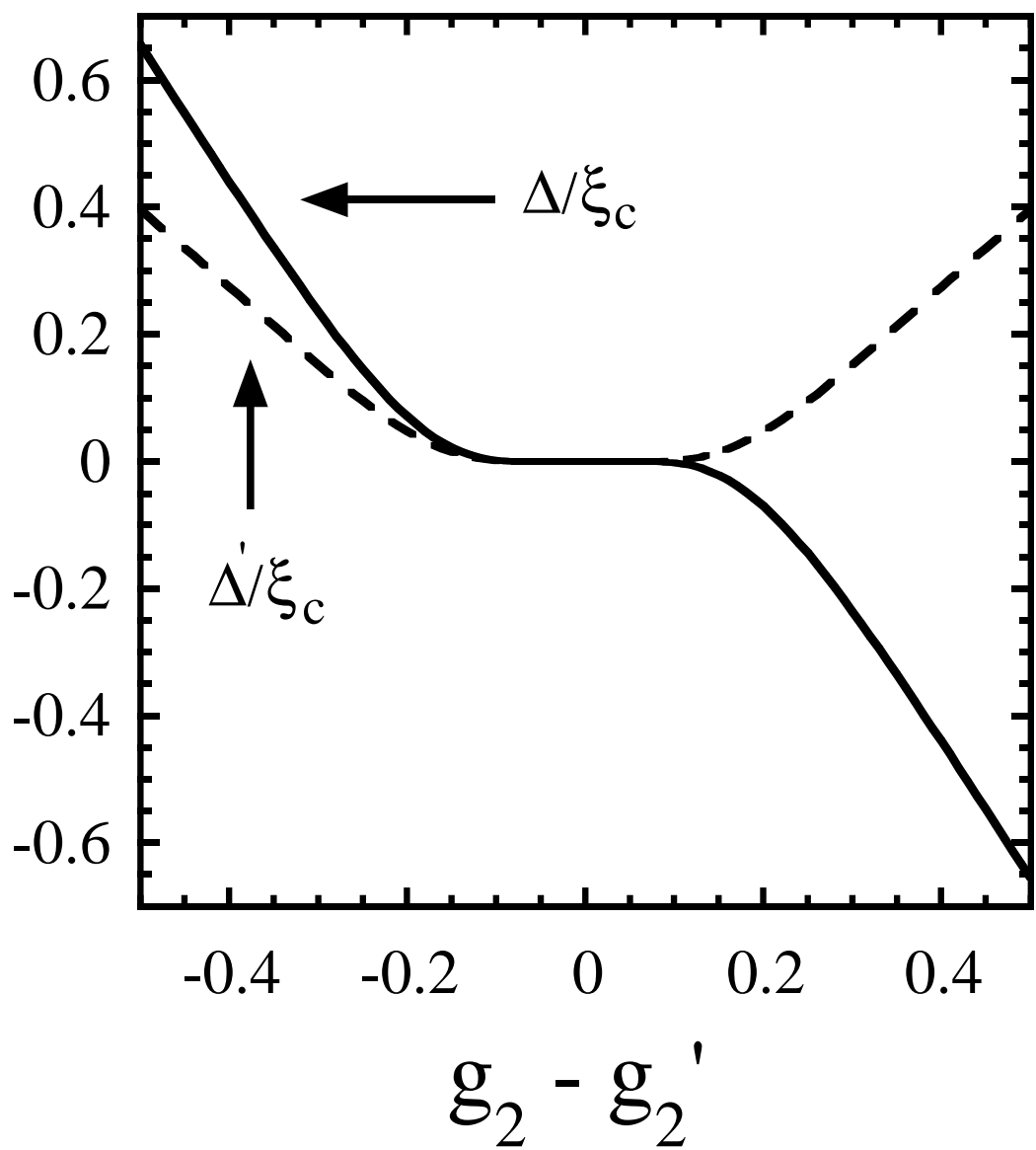
\begin{figure}
\caption
{
Solutions of the self-consistent equations, 
$\Delta / \xi_c$ and  $\Delta' / \xi_c $ as a function of $g_2-g_2'$, 
which are obtained from Eqs.(3.9) and (3.10).  
Note that $(-\Delta, -\Delta')$ is also the solution as well as $(\Delta, 
\Delta')$.
}
\label{fig:1}
\end{figure}
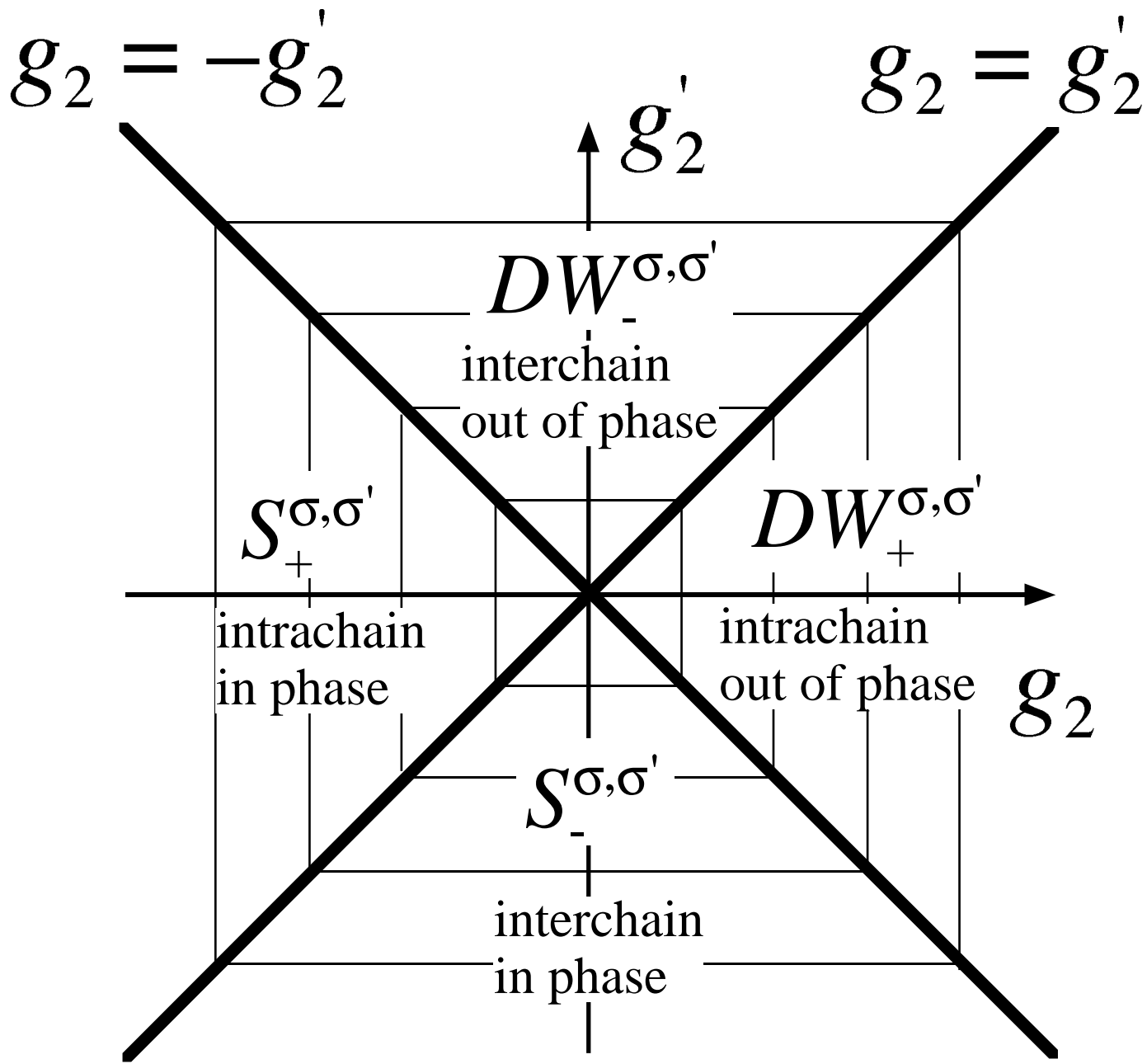
\begin{figure}
\caption
{
Phase diagram in the presence of the interchain hopping. 
Here $DW^{\p,\p'}_{-}$ ($DW^{\p,\p'}_{+}$) and  
$S^{\p,\p'}_{-}$ ($S^{\p,\p'}_{+}$) express 
density wave with interchain (intrachain) ordering with out of phase   
and 
superconductivity with interchain (intrachain) ordering with in phase, 
where $\p$ and $\p'$ express spin indices. 
 } 
\label{fig:2}
\end{figure}
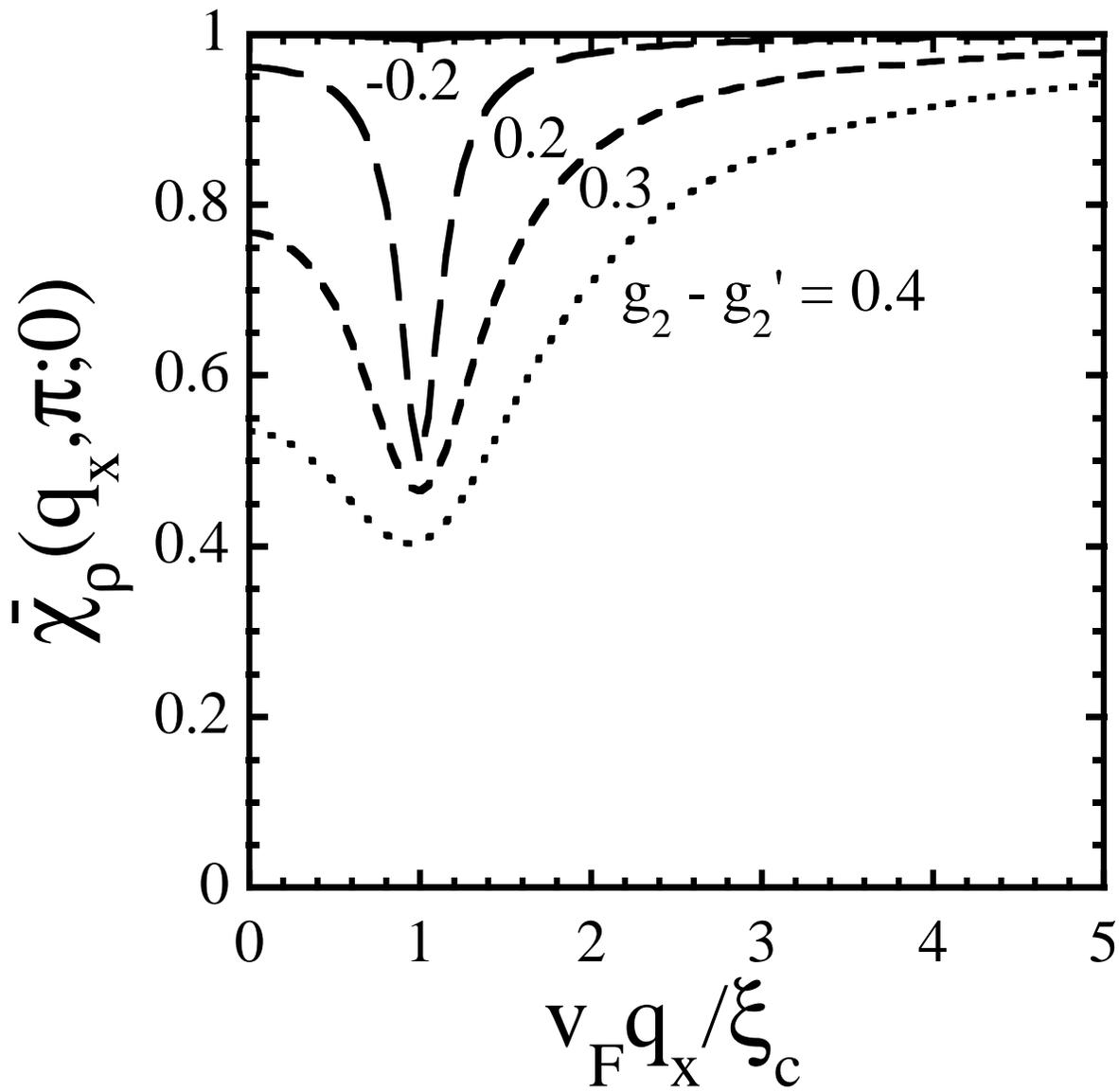
\begin{figure}
\caption
{
Normalized charge susceptibility, 
$ \bar \ch_\rho (q_x,\pi,0) ( \equiv  
 \Re \ch_\rho (q_x,\pi;0)/(2/\pi v_F)) $ 
 at $T=0$ as a 
function of $v_F q_x / \xi_c$ 
 in the case of  $g_2 - g_2' =$ -0.2, 0.2, 
0.3 and  0.4 where $\xi_c = 2t$.
} 
\label{fig:3}
\end{figure}
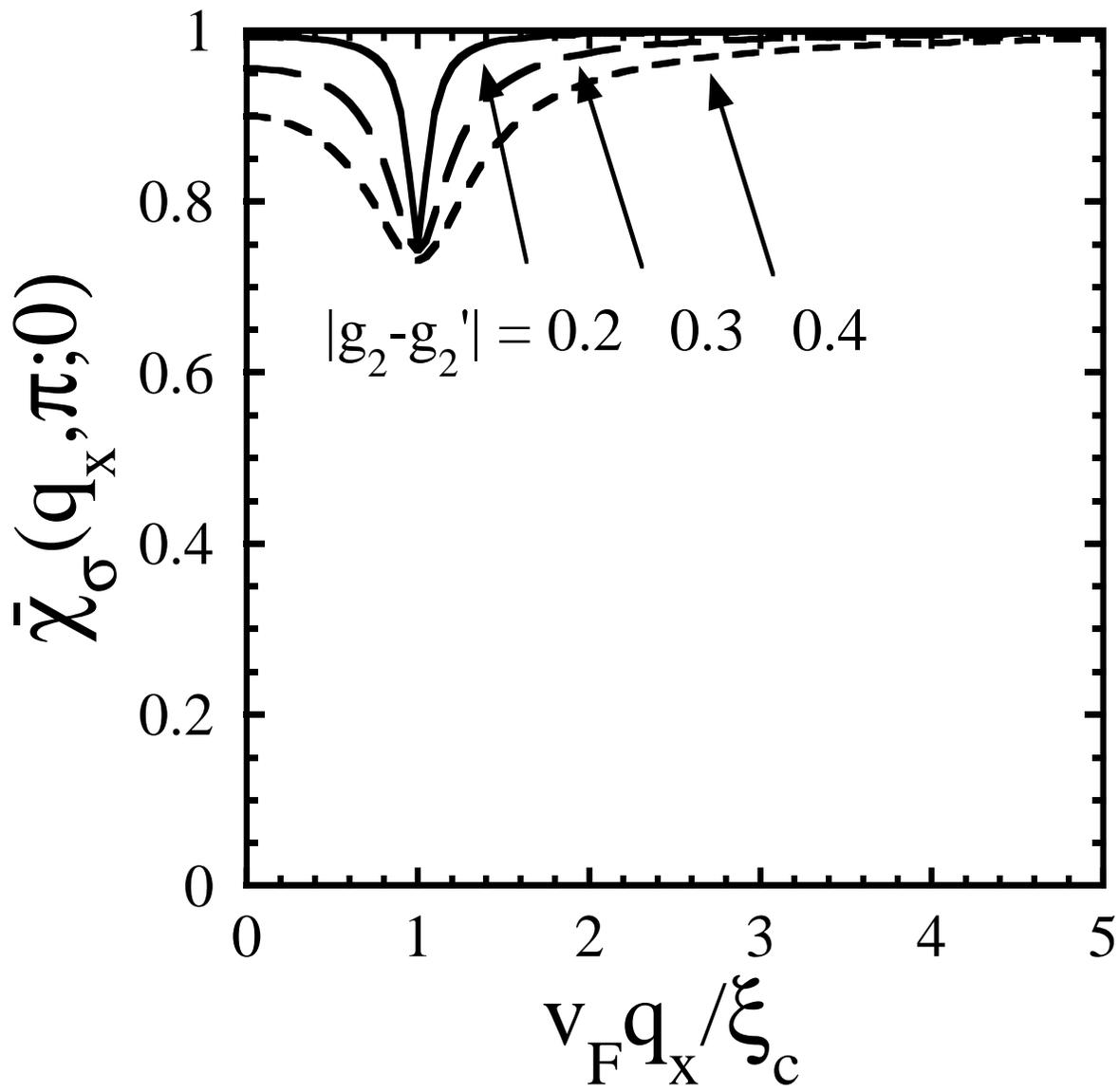
\begin{figure}
\caption
{
Normalized spin susceptibility, 
 $ \bar \ch_\p (q_x,\pi;0) 
 (\equiv \Re \ch_\p (q_x,\pi;0)/ (2/\pi v_F))$ 
 at $T=0$ as a function of $v_F q_x / \xi_c$ 
in the case of $|g_2 - g_2'| =$ 0.2, 0.3 and 0.4 where  
 $\xi_c = 2t$.
} 
\label{fig:4}
\end{figure}
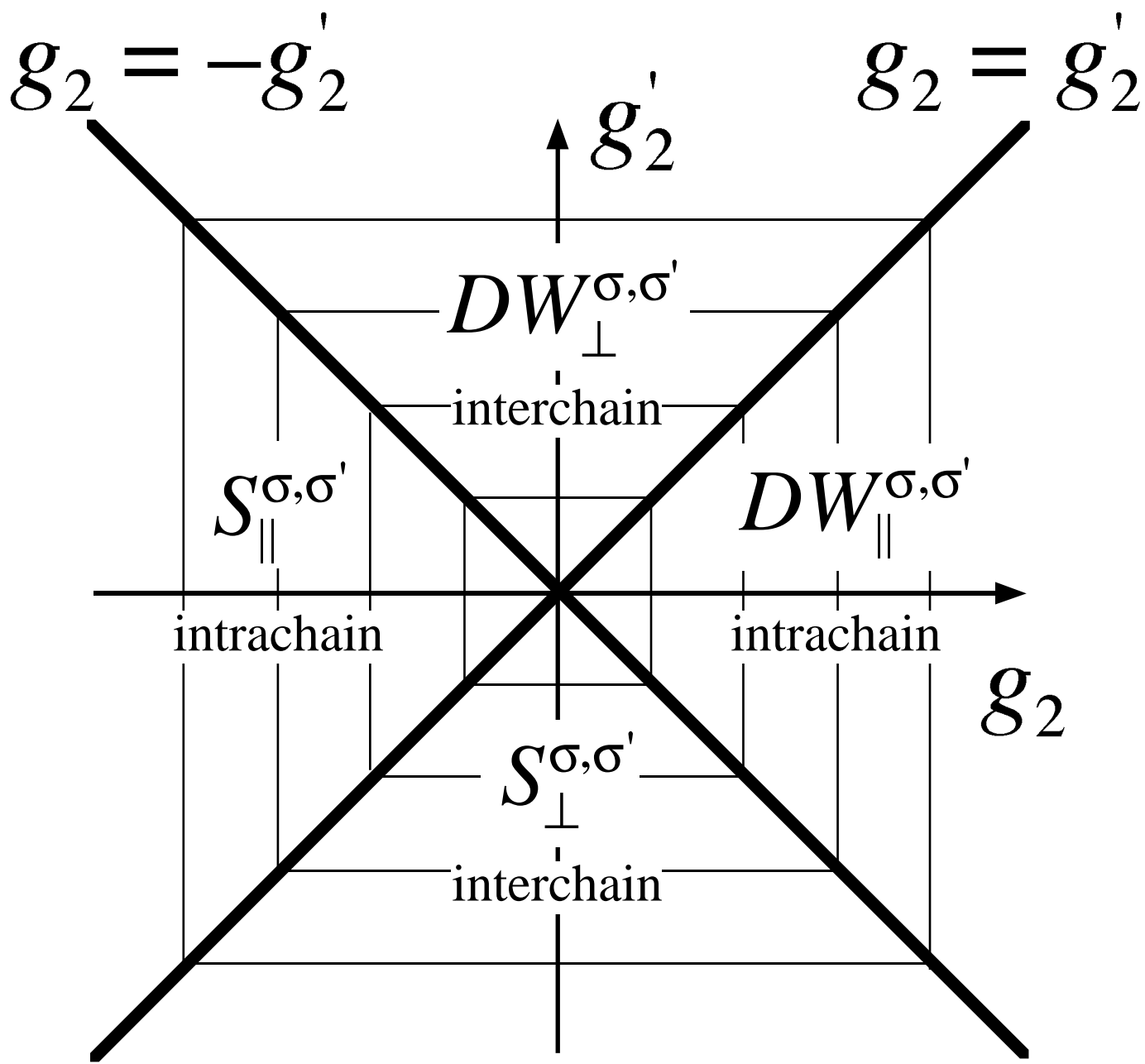
\begin{figure}
\caption
{
Phase diagram in the absence of the interchain hopping. 
Here $DW^{\p,\p'}_{\bot}$ ($DW^{\p,\p'}_{||}$) and  
$S^{\p,\p'}_{\bot}$ ($S^{\p,\p'}_{||}$) express 
density wave with interchain (intrachain) ordering and 
superconductivity with interchain (intrachain) ordering,   
where $\p$ and $\p'$ express spin indices. 
 } 
\label{fig:5}
\end{figure}


\begin{thebibliography}{99}
\bibitem{correspondence}
Correspondence between notations used by 
us and that by Schulz\cite{Schulz} are as follows, 
$\theta_+ = \sqrt{2} \phi_{\rho +}$,
$\theta_- = - \sqrt{2} \theta_{\rho +}$,
$\phi_+ = \sqrt{2} \phi_{\p +}$,
$\phi_- = - \sqrt{2} \theta_{\p +}$,
$\titheta_+ = \sqrt{2} \phi_{\rho -}$,
$\titheta_- = - \sqrt{2} \theta_{\rho -}$,
$\tiphi_+ = \sqrt{2} \phi_{\p -}$ and 
$\tiphi_- = - \sqrt{2} \theta_{\p -}$.
\end{thebibliography}
\end{document}